\documentclass[sigconf,nonacm]{acmart}

\usepackage{tikz}
\usepackage{enumitem}

\usepackage{graphicx} 
\usepackage{xspace}
\usepackage{listings}
\usepackage{verbatim}
\usepackage[normalem]{ulem}
\usepackage{csquotes}
\usepackage{amsmath}
\usepackage{amsfonts}
\usepackage{amsthm}
\usepackage{subcaption}
\usepackage{pifont}
\usepackage{multirow}
\usepackage{algorithm}
\usepackage{algpseudocode}
\usepackage{numprint}
\usepackage{hyperref}
\usepackage[capitalise, noabbrev]{cleveref}

\usepackage{bigints}
\usepackage{hhline}

\usepackage{makecell}
\usepackage{arydshln}

\newtheorem{theorem}{Theorem}
\newtheorem{proposition}{Proposition}

\usepackage{placeins}

\DeclareFontFamily{U}{mathb}{\hyphenchar\font45}
\DeclareFontShape{U}{mathb}{m}{n}{
<-6> mathb5 <6-7> mathb6 <7-8> mathb7
<8-9> mathb8 <9-10> mathb9
<10-12> mathb10 <12-> mathb12
}{}
\DeclareSymbolFont{mathb}{U}{mathb}{m}{n}
\DeclareMathSymbol{\llcurly}{\mathrel}{mathb}{"CE}
\DeclareMathSymbol{\ggcurly}{\mathrel}{mathb}{"CF}

\newcommand{\etal}{\textit{et al.}}
\newcommand{\cf}{cf.\xspace}

\newcommand{\WRTxs}{FR Txs}

\newcommand{\observer}{\ensuremath{o}\xspace}
\newcommand{\SObservers}{\ensuremath{\mathcal{O}}\xspace}
\newcommand{\nmbObservers}{\ensuremath{n}\xspace}
\newcommand{\nmbObserversSubset}{\ensuremath{m}\xspace}

\newcommand{\networksize}{\ensuremath{N}\xspace}

\newcommand{\transaction}{\ensuremath{t}\xspace}

\newcommand{\block}{\ensuremath{\mathcal{B}}\xspace}

\newcommand{\DistReception}{\ensuremath{\mathcal{D}^{(r)}}\xspace}

\newcommand{\DistReceptionLN}{\ensuremath{\mathcal{LN}}\xspace}
\newcommand{\timeDiff}{\ensuremath{\omega}\xspace}

\newcommand{\pr}[1]{\ensuremath{Pr\left[#1\right]}\xspace}

\newcommand{\receivedBefore}[2]{\ensuremath{#1 \prec\,#2}\xspace}
\newcommand{\receivedBeforeBy}[3]{\ensuremath{#1 \prec_{#3} #2}\xspace}

\newcommand{\receivedAfterBy}[3]{\ensuremath{#1 \succ_{#3} #2}\xspace}
\newcommand{\sentBefore}[2]{\ensuremath{#1 \llcurly #2}\xspace}

\newcommand{\cmark}{\ding{51}}
\newcommand{\xmark}{\ding{55}}

\newcommand{\timeVar}{\ensuremath{\tau}\xspace}
\newcommand{\sendTime}[1]{\ensuremath{\sigma\left(#1\right)}\xspace}

\newcommand{\censorTime}{\ensuremath{\tau_C}\xspace}

\newcommand{\PDF}[2]{\ensuremath{f_{#1}\!\!\left(#2\right)}\xspace}
\newcommand{\CDF}[2]{\ensuremath{F_{#1}\!\!\left(#2\right)}\xspace}

\newcommand{\probGMO}{\ensuremath{\mathbb{Q}}\xspace}
\newcommand{\PPV}{\ensuremath{\mathbb{P}}\xspace}

\newcommand{\funcZ}{\ensuremath{\mathcal{E}_{m/n}^{\receivedBefore{\transaction_1}{\transaction_2}}}\xspace}

\newcommand{\subheading}[1]{\vspace{0.25 em}\noindent \textbf{#1}}

\begin{document}

\title{On the Effectiveness of Mempool-based Transaction Auditing}

\author{Jannik Albrecht}
\email{jannik.albrecht@ruhr-uni-bochum.de}
\affiliation{%
  \institution{Ruhr University Bochum}
  \city{Bochum}
  \country{Germany}
}
\author{Ghassan Karame}
\email{ghassan@karame.org}
\affiliation{%
  \institution{Ruhr University Bochum}
  \city{Bochum}
  \country{Germany}
}

\begin{abstract}
While the literature features a number of proposals to defend against transaction manipulation attacks, existing proposals are still not integrated within large blockchains, such as Bitcoin, Ethereum, and Cardano. Instead, the user community opted to rely on more practical but ad-hoc solutions (such as Mempool.space) that aim at \emph{detecting} censorship and transaction displacement attacks by auditing discrepancies in the mempools of so-called observers.

In this paper, we precisely analyze, for the first time, the interplay between mempool auditing and the ability to detect censorship and transaction displacement attacks by malicious miners in Bitcoin and Ethereum. Our analysis shows that mempool auditing can result in mis-accusations against miners with a probability larger than 25\% in some settings. On a positive note, however, we show that mempool auditing schemes can successfully audit the execution of any two transactions (with an overwhelming probability of 99.9\%) if they are consistently received by all observers and sent at least 30 seconds apart from each other. As a direct consequence, our findings show, for the first time, that batch-order fair-ordering schemes can offer only strong fairness guarantees for a limited subset of transactions in real-world deployments.
\end{abstract}

\maketitle

\section{Introduction}

Decentralized platforms continue to gain considerable traction as they eliminate the need for intermediaries. 
Here, transactions are first propagated in the P2P layer across blockchain nodes; correct and verified transactions are buffered by these nodes in a local data structure, the \emph{mempool}. Transactions in the mempool are then confirmed by miners---specific nodes that are rewarded for their effort---into so-called blocks. 
While this establishes a clear temporal order among transactions confirmed in different blocks, a malicious miner can censor, displace, or re-order unconfirmed transactions in its local mempool to increase its advantage in the system (e.g., by prioritizing transactions that pay higher fees or by dropping or censoring transactions).

Given that these attacks exploit intrinsic properties of existing public blockchains, finding practical clean-slate solutions against them emerges as a significant research challenge that received considerable attention from both industry and academia~\cite{DBLP:journals/corr/abs-1904-05234,DBLP:conf/fc/EskandariMC19,DBLP:conf/fc/QinZLG21,DBLP:conf/fc/McCorryHM18}. 
Existing remedies consist of preventing miners from enforcing a particular order when confirming blocks~\cite{DBLP:conf/asiapkc/KelkarDK22,DBLP:conf/ccs/KelkarDLJK23,DBLP:conf/aft/Kursawe20}. Unfortunately, these approaches either require considerable changes to the underlying blockchain protocols~\cite{DBLP:conf/aft/Kursawe20} or \emph{require consistent transaction ordering---an assumption that has not been empirically evaluated---to provide so-called batch-fair ordering.} Due to these limitations, such solutions are still not integrated into large blockchains, such as Bitcoin, Ethereum, Cardano, Litecoin, etc.

\begin{figure}[tbp]
    \centering
    \begin{subfigure}[t]{0.34\columnwidth}
	\includegraphics[width=\linewidth]{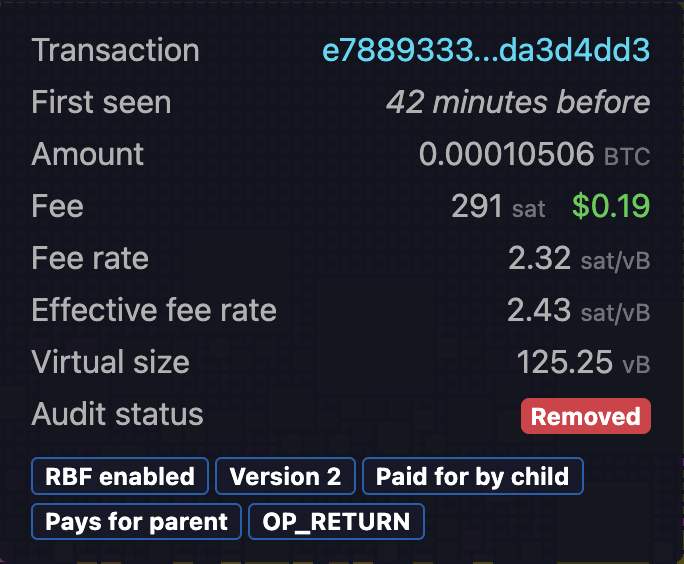}
	\caption{A potentially censored transaction.}
	\label{fig:mempool_block_health}
    \end{subfigure}
    \hspace{0.35em}
    \begin{subfigure}[t]{0.625\columnwidth}
        \includegraphics[width=\linewidth]{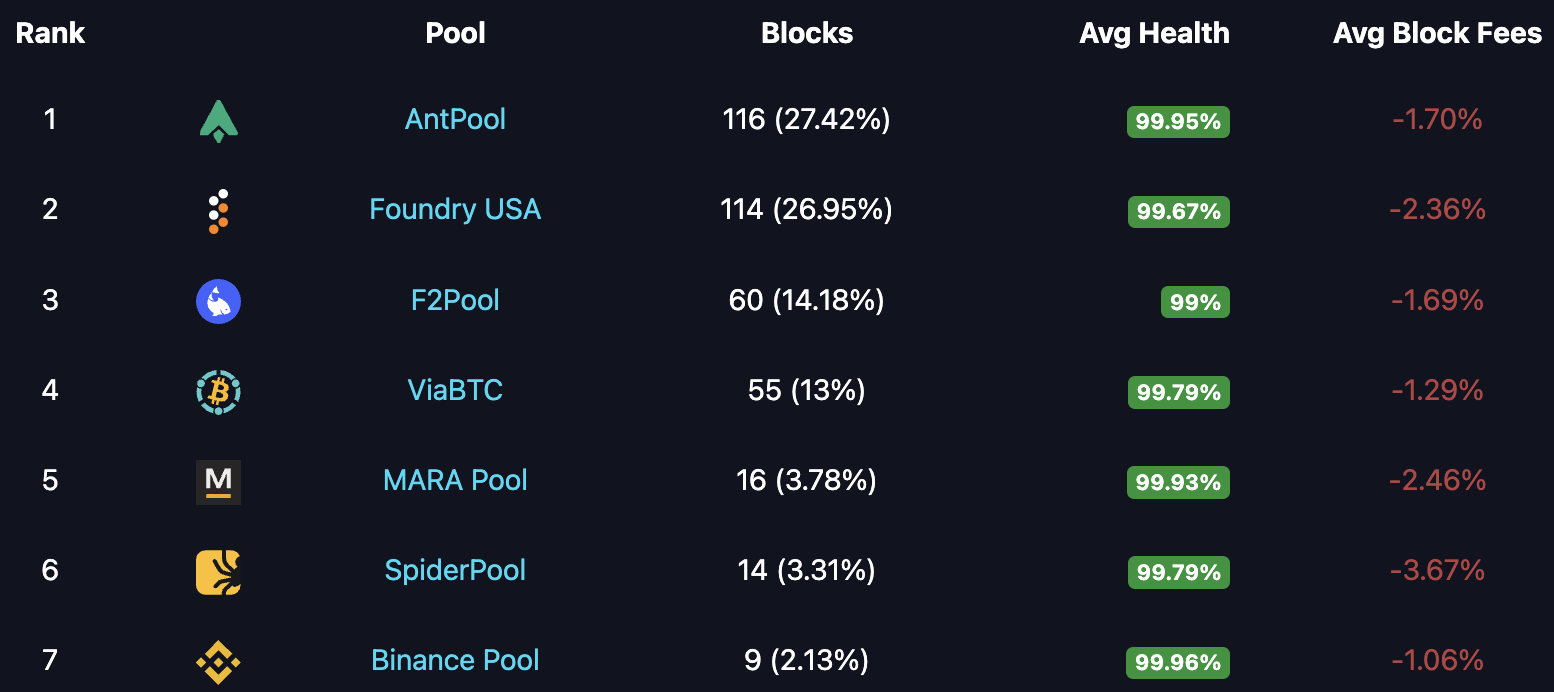}
        \caption{Block health ranking of the leading Bitcoin pools.}
        \label{fig:pools_ranking}
    \end{subfigure}
    \caption{An excerpt of the block audit by the Mempool Open Source Project~\cite{mempool_space}. Figure (a) shows an audit of a block's \emph{removed}, i.e., potentially censored, transaction. 
    Figure (b) shows Mempool.space's ranking of the leading Bitcoin pools in terms of block health.} 
    \label{fig:mempool_block_audit}
\end{figure}

Instead, the community opted to rely on practical but ad-hoc solutions~\cite{mempool_space, litecoin_space, blockstream, btcscan, etherscan, tokenview, blockchain_com_btc, blockchain_com_eth} that aim at \emph{detecting} censorship and transaction displacement attacks by untrusted miners by observing discrepancies in the mempools of so-called observer nodes. More precisely, these solutions aim to identify whether the transactions received by independent network observers match those confirmed in blocks output by the miners. We dub this approach \emph{mempool-based auditing}. For instance, Mempool.space~\cite{mempool_space} and Litecoin.space~\cite{litecoin_space} compute the so-called ``block health'' output by each mining pool: this metric measures the difference between the expected transactions in a block and the actual block output by a given mining pool. This can strongly influence users and deter malicious miners, as a few mining pools control mining/leader election in many popular cryptocurrencies~\cite{hashrate_distribution, Eth_block_producers} (see Figure~\ref{fig:pools_ranking}).

In this paper, we set forth to precisely analyze, for the first time, the provisions of mempool auditing in detecting censorship and transaction displacement attacks by miners. Our analysis explores the precise guarantees that such \emph{mempool-based} detection approaches can give in realistic deployments, such as the current Bitcoin and Ethereum ecosystems. More precisely, we are interested in answering fundamental questions such as: \emph{what is the precise probability that any two transactions sent in order will be observed in the same order by up to $n$ independent observers? and what are the implications on the efficiency of batch--fair ordering schemes}, \emph{Under what specific conditions can we confidently conclude that no censoring or transaction displacement attack has occurred?} This is particularly challenging to model since the local views of honest observers depend on factors such as connectivity, geographical distribution, and other network peculiarities, which drastically affect the guarantees exhibited by such auditing approaches. 

More specifically, we make the following contributions in this work:\footnote{The code and data required to reproduce our results are available at \cite{artifact}.}

\begin{description}[leftmargin=0.5 cm]
\setlength\itemsep{0em}
    \item[Network model: ] We devise a realistic network model to demarcate the limits of (im-)possibility, and to capture all the relevant notions for mempool-based detection approaches (\cf~Section~\ref{sec:network_model}). 
    \item[Large-scale measurements: ] 
    We validate our analytical model using large-scale measurements in Bitcoin and Ethereum.  
    The combined outcomes of our formal and empirical analyses provide fundamental insights toward understanding the impact of the network layer on the ability to detect censorship and displacement attacks in the wild. 
    \item[Effectiveness of mempool-auditing: ] 
    Using both analytical and empirical methods, we show that mempool auditing exhibits limited generalizability and can lead to serious misaccusations due to inherent inconsistencies. More specifically, our findings reveal that a transaction that is sent more than 30 seconds earlier than another one is (almost) guaranteed to be received before the latter (\cf~Section~\ref{sec:network_model}). 
    Our results further suggest that mempool auditing can mislabel honest miners as ``malicious'' with a probability larger than 25\% in some settings (\cf~Section~\ref{sec:eth_eval}). 
    \item[First analysis of batch-fair ordering schemes: ] Our results also have direct implications for evaluating the effectiveness of fair-ordering schemes. Specifically, we demonstrate that the key assumption underlying batch-fair ordering schemes---namely, that validators receive a consistent transaction order---is rarely satisfied in practice (\cf~Section~\ref{sec:interpretation_of_findings}).
\end{description}

\begin{figure}[tbp]
    \centering
    \begin{subfigure}[t]{0.45\columnwidth}
	\includegraphics[width=0.97\linewidth]{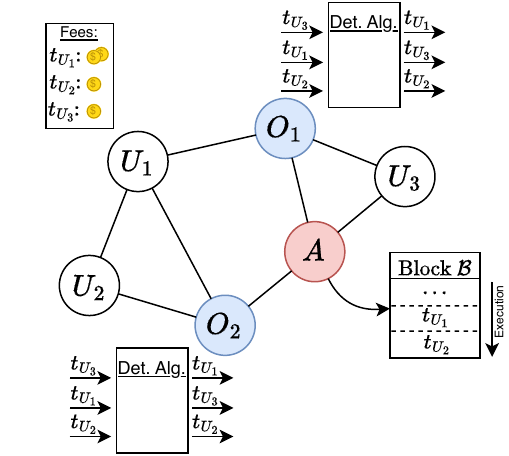}
	\caption{Censoring attack. }
	\label{fig:example_censoring}
    \end{subfigure}
    \hspace{1.0em} 
    \begin{subfigure}[t]{0.45\columnwidth}
	\includegraphics[width=0.97\linewidth]{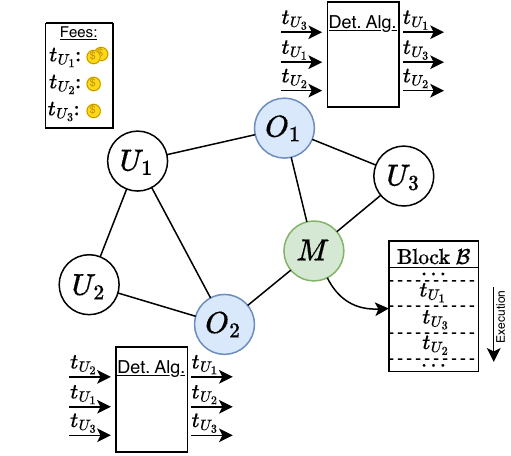}
	\caption{Observers in dispute.}
	\label{fig:example_dispute}
    \end{subfigure}
    \caption{Examples showing adversary $A$ executing displacement attacks: In Figure (a), adversary $A$ censors transaction $t_{U_3}$ by omitting it from block $\mathcal{B}$, while two observers $O_1$ and $O_2$ expect $t_{U_3}$ to be scheduled before $t_{U_2}$. Figure (b) shows a benign example, where both observers are in dispute on the rightful order of $t_{U_2}$ and $t_{U_3}$.  } 
    \label{fig:example_displacement_attacks}
\end{figure}


\section{Problem Statement} \label{sec:problem_statement}

While blocks impose a strict order on confirmed transactions, unconfirmed ones can be reordered by miners to gain advantage---for example, by prioritizing higher-fee transactions. Such manipulations are commonly referred to as front-running attacks, where adversaries observe pending transactions and attempt to have their own confirmed first. Miners may further delay transactions indefinitely, enabling censorship or targeted exclusion. These attacks are prevalent in practice, with consequences ranging from financial losses (e.g., arbitrage or stolen secrets) to censorship. For instance, Qin \etal~\cite{DBLP:conf/sp/QinZG22} report losses of \$540.54M over 32 months (Dec 2018–Aug 2021) due to arbitrage attacks. In the sequel, we refer to all such manipulations as ``transaction displacement attacks''\footnote{Terminology in the literature is inconsistent: ``front-running'' may denote both the broader class of order-manipulation attacks and the specific insertion of malicious transactions.}.

Unfortunately, existing solutions to deter displacement attacks are impractical for existing blockchains. Namely, they either cannot support real-time detection/prevention or are not generically applicable to different blockchains/applications (see Section~\ref{sec:apdx_rel_work} in the Appendix for additional details). This gave rise to ad-hoc solutions~\cite{mempool_space, litecoin_space, blockstream, btcscan, etherscan, tokenview, blockchain_com_btc, blockchain_com_eth} that aim at \emph{detecting} censorship and transaction displacement attacks by untrusted miners by observing discrepancies in the mempools of so-called observer nodes.

\subheading{System Model.}
We consider a blockchain platform built atop a peer-to-peer (P2P) network connecting \networksize nodes. Transactions are propagated among those nodes over the P2P layer. Each node stores the valid transactions it receives in its local mempool, a local data structure that maintains an ordered list of valid transactions. We assume that all honest nodes order transactions in their mempool based on their order of arrival. In practice, however, rational nodes tend to prioritize transactions that yield larger rewards, i.e., larger fees per weight ratio, to incorporate those in a block. Hence, we capture this strategy of rational miners\footnote{Note that while mechanisms like Ethereum’s proposer-builder separation (PBS) divide block generation between multiple parties, such as proposers and builders, entities that are responsible for assembling blocks using their own local transaction views remain vulnerable to ordering inconsistencies.} 
by assuming the usage of a public deterministic algorithm (\cf Det.~Alg.\ in \cref{fig:example_displacement_attacks}) to schedule the transactions' execution based on their mempool order, their fees, their amounts, etc. This aligns with the behavior of widely used blockchain clients, such as bitcoind~\cite{bitcoind_repo} and geth~\cite{geth_repo}, which order transactions by reception time and then apply public and deterministic algorithms, such as fee-based sorting.
Similar to~\cite{DBLP:conf/uss/TorresCS21,DBLP:conf/sp/QinZG22,DBLP:journals/corr/abs-2106-07371}, we assume that any deviation from this common strategy---such as the use of probabilistic or proprietary ordering---is an instance of censorship or misbehavior. 

Given that the algorithm is public, observing either the input, e.g., the reception order, or the output, i.e., the sorted mempool, provides equivalent information for our analysis. In other words, given some input to the algorithm, the transactions' order in mempool can be easily predicted for \emph{honest nodes}. As soon as enough transactions have been received to fill a block, all transactions received at a later time are scheduled by miners to be confirmed in a subsequent block. We denote this time when enough transactions have been received to fill a block as \censorTime (\cf~Section~\ref{sec:DaB}). 

We assume that rational miners aim to increase their profit in the system while avoiding detection (i.e., to ensure that their reputation is not damaged (\cf \cref{fig:pools_ranking})).  
\emph{Specifically, we focus on the ability of malicious miners to select transactions from their local mempool and freely order them} to increase their advantage in the system. We do not consider the (similar) impact that non-miners can achieve by rushing their transactions in the system; indeed, while the impact of such attacks is detrimental, miners eventually have the ultimate decision in determining the eventual order of transactions and could correct this behavior.

\subheading{Mempool-auditing.} Several solutions in the community, such as mempool.space~\cite{mempool_space} and Etherscan~\cite{etherscan}, and some fair-ordering schemes, like Themis~\cite{DBLP:conf/ccs/KelkarDLJK23}, rely on the core assumption that, since all nodes use a deterministic algorithm to sort their local mempool, the various validators should exhibit a similar order of transactions (assuming ideal network functionality). 
This core assumption allows for deterrence of misbehavior by \emph{detecting} malicious miners who report a different transaction order in their confirmed blocks than the one witnessed by a ``trusted'' set of nodes, called observers, which are commissioned by a service provider to audit block/transaction ordering.
An overview of existing Bitcoin and Ethereum blockchain detection tools is summarized in \cref{tab:detection_tools}. These tools mostly monitor the reception of transactions and blocks and provide some web interface that logs pending transactions and their reception times (\cf~\cite{blockstream, btcscan, etherscan, tokenview, blockchain_com_btc, blockchain_com_eth}). These logs can be utilized to (1) validate the reception of a transaction and (2) estimate the time for a transaction to be confirmed in a block. 
Unexpected deviations from the presumed confirmation schedule indicate a potential misbehavior.
We refer to this approach as mempool-based transaction auditing and depict it in \cref{fig:example_displacement_attacks}. 
Here, \cref{fig:example_censoring} shows two observers $O_1, O_2$ that witnessed three transactions $\transaction_{U_1}$, $\transaction_{U_2}$, and $\transaction_{U_3}$. Both $O_1$ and $O_2$ receive $\transaction_{U_3}$ before $\transaction_{U_1}$ and $\transaction_{U_2}$ (denoted as $\receivedBefore{\transaction_{U_3}}{\transaction_{U_1}}$, $\receivedBefore{\transaction_{U_3}}{\transaction_{U_2}}$), while the adversary $A$ proposes a block $\block$ that executes both $\transaction_{U_1}$ and $\transaction_{U_2}$ but omits $\transaction_{U_3}$.

Unfortunately, while such auditing schemes can emerge as a deterrent against front-running attacks, they can only offer probabilistic guarantees. Namely, it is possible that observers do not always witness a similar order to that of a given miner due to delays, network topology, or other aspects that impact network connectivity. We illustrate this by means of an example in Figure~\ref{fig:example_dispute}. Here, honest observers $O_1$ and $O_2$ receive transactions $\transaction_{U_1}$, $\transaction_{U_2}$, and $\transaction_{U_3}$ in different orders due to network jitter.

\subheading{Research Questions: }
In this paper, we set forth to explore the precise guarantees that mempool-based transaction auditing approaches can give in blockchain deployments. More specifically, we aim to answer the following fundamental research questions:

\begin{description}
\setlength\itemsep{0em}
    \item[RQ1] What is the probability that up to $n$ independent observers observe the same set of transactions in Bitcoin and Ethereum?
    \item[RQ2] What specific conditions must be satisfied for mempool auditing schemes to conclude with high confidence that no transaction displacement \emph{within blocks} has occurred?
    \item[RQ3] What specific conditions must be met for mempool auditing schemes to reliably determine that no transaction displacement \emph{across different blocks} has taken place? Similarly, when can mempool auditing schemes conclude with high certainty that no censoring has occurred?
    \item[RQ4] Based on our analysis, how do threshold fair-ordering schemes that rely on consistent transaction orderings among validators---such as Themis~\cite{DBLP:conf/ccs/KelkarDLJK23}—perform compared to traditional fair-ordering approaches like Pompe~\cite{DBLP:conf/osdi/ZhangSCZA20} and Wendy~\cite{DBLP:conf/aft/Kursawe20}? 
\end{description}

\section{Mempool-based Network Auditing}\label{sec:network_model}

\subsection{Network Model}\label{sec:network_behavior}

In what follows, we determine the probability that a share of \nmbObserversSubset out of the total \nmbObservers observers agree on two transactions' order. First, we determine the probability that a single observer receives two transactions sent \timeDiff apart in a certain order (\cf~Equation~(\ref{eq:prob_head_int})). Then, we generalize this probability to account for multiple observers. 

We start by computing the probability that an observer $\observer$ receives a transaction $\transaction_1$ before another transaction $\transaction_2$ that was sent \timeDiff time after $\transaction_1$.
For this, we introduce $\sendTime{\transaction}$ denoting the time when a transaction $\transaction$ is sent and $\timeVar_{\observer}(\transaction)$ as the time it takes to deliver transaction \transaction from its sender to observer \observer. 
Here, $\timeVar_{\observer}(\transaction)$ is a random variable that follows the distribution\footnote{We assume that \DistReception is agnostic to a specific network topology or configuration.} of propagation times $\timeVar_{\observer}(\transaction) \sim \DistReception$. 
Then, given $\transaction_1, \transaction_2$ sent at time $\sendTime{\transaction_1}, \sendTime{\transaction_2}$ with $\sendTime{\transaction_2} = \sendTime{\transaction_1} + \timeDiff$ and received after $\timeVar_{\observer}(\transaction_1), \timeVar_{\observer}(\transaction_2)$ time by observer $\observer$, the probability that \observer receives $\transaction_1$ before $\transaction_2$, i.e., $\pr{\receivedBeforeBy{\transaction_1}{\transaction_2}{\observer}}$, is equal to:
\begin{align}
    \pr{\receivedBeforeBy{\transaction_1}{\transaction_2}{\observer}} = \int_0^\infty \CDF{\timeVar_{\observer}(\transaction_1)}{\timeDiff + \iota} \PDF{\timeVar_{\observer}(\transaction_2)}{\iota} d\iota.  \label{eq:prob_head_int}
\end{align}
Here, \PDF{\timeVar_{\observer}(\transaction)}{T} denotes the probability density function (PDF) of the distribution of propagation times \DistReception. \CDF{\timeVar_{\observer}(\transaction)}{T} denotes the cumulative distribution function (CDF) of \DistReception, i.e., the probability that $\timeVar_{\observer}(\transaction)$ is less than (or equal to) a value $T$.
Further, it holds $\pr{\receivedAfterBy{\transaction_1}{\transaction_2}{\observer}} = 1 - \pr{\receivedBeforeBy{\transaction_1}{\transaction_2}{\observer}}$.

\begin{figure}[tb]
    \centering
    \begin{subfigure}[t]{0.49\columnwidth}
        \centering
	    \includegraphics[width=1\linewidth]{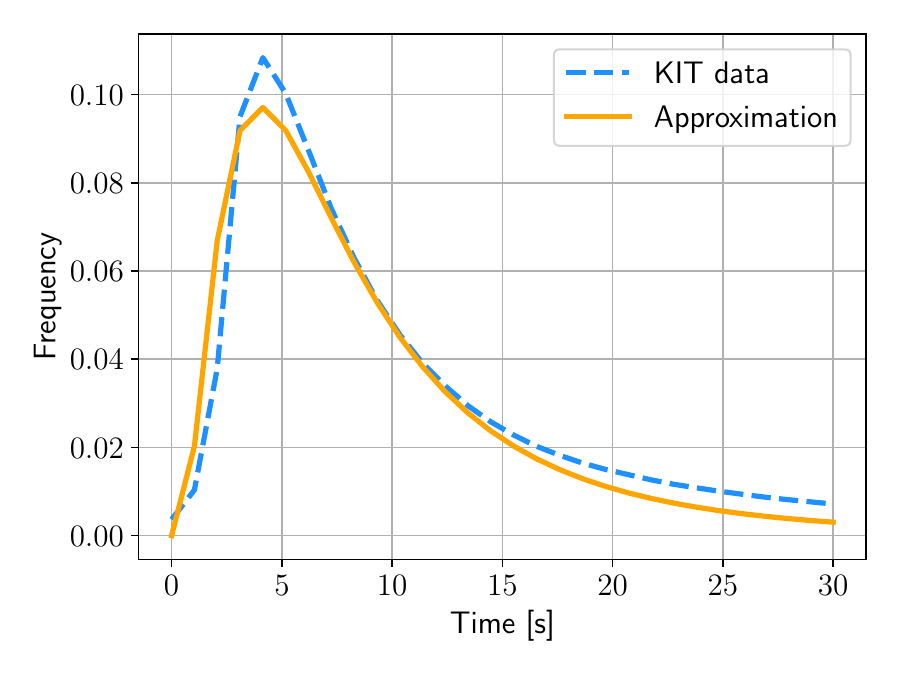}
	    \caption{PDF of the transactions' transmission times.}
	    \label{fig:KIT_PDF}
    \end{subfigure} 
    \begin{subfigure}[t]{0.49\columnwidth}
        \centering
        \includegraphics[width=1\linewidth]{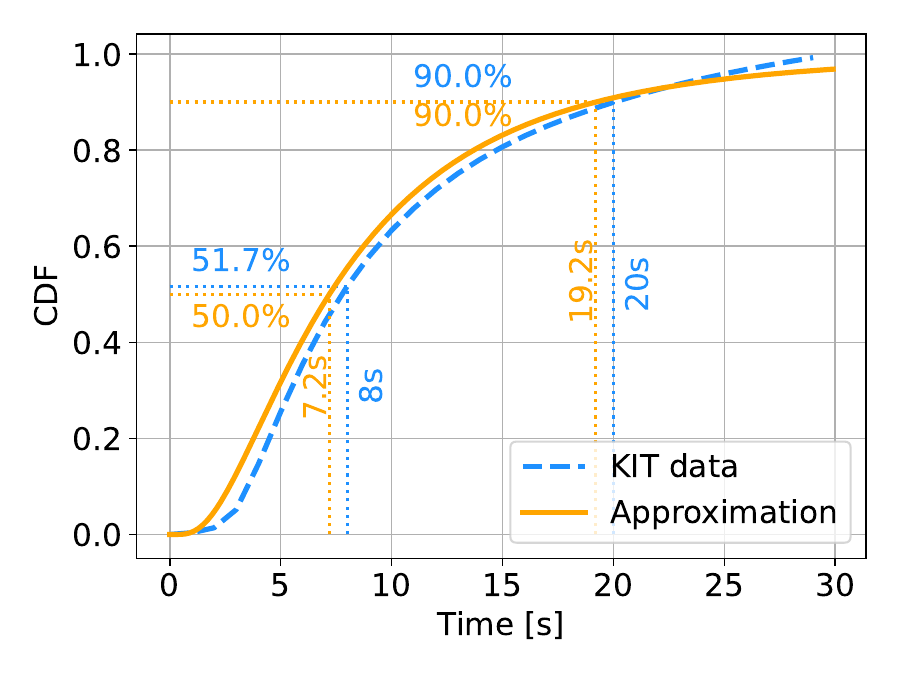}
        \caption{CDF of the transactions' transmission times. }
	    \label{fig:KIT_CDF}
    \end{subfigure}
    \caption{The PDF/CDF of transaction transmission times based on the KIT dataset (dashed) and its log-normal approximation (solid).} 
	\label{fig:KIT_dist}
\end{figure}

The probability $\pr{\receivedBeforeBy{\transaction_1}{\transaction_2}{\observer}}$ is parameterized by \timeDiff, i.e., the time difference between sending the two transactions $\transaction_1$ and $\transaction_2$. 
Evaluating it with a small value \timeDiff compares the probabilities of two (almost) equiprobable events. For $\timeDiff=0$, $\pr{\receivedBeforeBy{\transaction_1}{\transaction_2}{\observer}}$ yields a probability of 50\%. Moreover, increasing \timeDiff causes the integral shown in Equation~(\ref{eq:prob_head_int}) to monotonically increase.

Next, we leverage Equation~(\ref{eq:prob_head_int}) to prove Theorem~\ref{th:generalized_multiple_observers}, which generalizes the probability to account for multiple observers. 

That is, we determine the probability that \nmbObserversSubset out of \nmbObservers observers ($\nmbObserversSubset \leq \nmbObservers$) witness the same order among two transactions. Ultimately, this probability can be leveraged to assess the applicability of an observer-based auditing scheme to detect re-ordering attacks. 
Here, we assume that the propagation times $\timeVar_{\observer_i}(\transaction)$ and $\timeVar_{\observer_j}(\transaction)$, i.e., the times it takes for a single transaction $\transaction$ to reach two different observers $\observer_i$ and $\observer_j$, are independent random variables for any pair of distinct observers $\observer_i \neq \observer_j$. 
This assumption is based on (1) the observers being randomly positioned within the network, (2) senders are likewise expected to be randomly distributed, and (3) observers are not directly connected, preventing direct snapshot correlations. 
We further confirm this assumption by empirical measurements  (\cf~Section~\ref{sec:bc_eval}). We now introduce our first main theorem.

\begin{theorem} \label{th:generalized_multiple_observers}
    Given are two transactions $\transaction_1, \transaction_2$ with $\transaction_2$ sent \timeDiff time after $\transaction_1$ and \nmbObservers observers logging network events. 
    Then, the probability that exactly \nmbObserversSubset out of \nmbObservers observers ($\nmbObserversSubset \leq \nmbObservers$) receive transaction $\transaction_1$ before $\transaction_2$ is given by:
    \begin{align}
        \probGMO_{\nmbObservers}(\nmbObserversSubset) = & Pr\Big[\ \overbrace{\nmbObserversSubset = \bigg|\Bigl\{\observer\in\SObservers\ \big|\ (\receivedBeforeBy{\transaction_1}{\transaction_2}{\observer}) \Bigr\}\bigg| }^{\funcZ} \ \Big] \\ 
        &= \binom{\nmbObservers}{\nmbObserversSubset} \Biggl( \left( \int_0^\infty \CDF{\timeVar_{\observer}(\transaction_1)}{\timeDiff + \iota} \PDF{\timeVar_{\observer}(\transaction_2)}{\iota} d\iota \right)^{\nmbObserversSubset}  \nonumber \\  
        & \qquad \cdot \left( 1 - \int_0^\infty \CDF{\timeVar_{\observer}(\transaction_1)}{\timeDiff + \iota} \PDF{\timeVar_{\observer}(\transaction_2)}{\iota} d\iota \right)^{\nmbObservers-\nmbObserversSubset} \Biggr) \nonumber
    \end{align}
\end{theorem}

Here, the term $\funcZ$ denotes the event that exactly \nmbObserversSubset out of \nmbObservers observers receive transaction $\transaction_1$ before transaction $\transaction_2$. It is defined:
\begin{equation}
    \funcZ := \quad \nmbObserversSubset = \bigg|\Bigl\{\observer\in\SObservers\ \big|\ (\receivedBeforeBy{\transaction_1}{\transaction_2}{\observer}) \Bigr\}\bigg|.
\end{equation}

Function $\probGMO_{\nmbObservers}(\nmbObserversSubset)$ relates to the probability mass function of the binomial distribution. That is, we model the event $\pr{\receivedBeforeBy{\transaction_1}{\transaction_2}{\observer}}$ for any observer $\observer$ as a Bernoulli trial. Additionally, we consider any of these Bernoulli trials successful when $\transaction_1$ is received before $\transaction_2$ by the respective observer. Then, $\probGMO_{\nmbObservers}(\nmbObserversSubset)$ determines the probability that exactly \nmbObserversSubset out of \nmbObservers Bernoulli trials are successful.  
Here, $\probGMO_{\nmbObservers}(\nmbObserversSubset)$ is parameterized by three parameters. These parameters consist of (i) the time difference in sending $\transaction_1$ and $\transaction_2$ \timeDiff, (ii) the number of observers \nmbObservers, and (iii) the size of the observers' subset that witness the same order of $\transaction_1$ and $\transaction_2$ \nmbObserversSubset. 
Due to space limitations, the proof of Theorem~\ref{th:generalized_multiple_observers} can be found in Section~\ref{sec:proofs}.
Note that $\probGMO_{\nmbObservers}(\nmbObserversSubset)$ monotonically increases with increasing \timeDiff for fixed values \nmbObserversSubset and \nmbObservers. 
We evaluate $\probGMO_{\nmbObservers}(\nmbObserversSubset)$
in Table~\ref{tab:validate_Q_numerically_smallN} and Table~\ref{tab:validate_Q_numerically_largeN} based on real-world measurements in the Bitcoin blockchain (\cf~Section~\ref{sec:kit_data}).

\subsection{Witnessing In-order Events} \label{sec:anomaly_implications}

We now determine the probability that a transaction $\transaction_1$ has been sent before $\transaction_2$ conditioned on the event that \nmbObserversSubset out of \nmbObservers observers reported the reception of $\transaction_1$ before $\transaction_2$. This basically allows us to assess the effectiveness of mempool-based auditing schemes. 

\begin{theorem}[] \label{th:predictive_precision}
    Given two transactions $\transaction_1, \transaction_2$ and \nmbObservers observers, the probability that $\transaction_1$ has been sent before $\transaction_2$ given that \nmbObserversSubset out of \nmbObservers observers received $\transaction_1$ before $\transaction_2$ is:
    \begin{align}
        \PPV_{\nmbObservers}(\nmbObserversSubset) &= \pr{\sentBefore{\transaction_1}{\transaction_2} \ \big| \ \funcZ }   \nonumber \\ 
        &= \bigintss_{0}^{\infty} \left( \frac{ \pr{\funcZ \ \Big|\ \timeDiff} \cdot \pr{\timeDiff} }{ \bigintss_{-\infty}^{\infty} \pr{\funcZ \ \Big|\ \widehat{\timeDiff}} \cdot \pr{\widehat{\timeDiff}} \; d\widehat{\timeDiff} } \right) \; d\timeDiff 
    \end{align}
\end{theorem}

$\PPV_{\nmbObservers}(\nmbObserversSubset)$ estimates the probability that the original sending order of transactions is indeed similar to observers' local event logs. 
The more observers agree on either the event $\receivedBefore{\transaction_1}{\transaction_2}$ or the event $\receivedBefore{\transaction_2}{\transaction_1}$, the more likely is \timeDiff to be greater or lower than 0. Further, a larger value \nmbObservers, i.e., more observers in the network, allows more confident predictions. That is, the prediction based on a few observers' observations, e.g., $\nmbObservers \leq 3$ is more error-prone than a prediction that is based on many observing auditors, e.g., $\nmbObservers \geq 10$. 
We evaluate $\PPV_{\nmbObservers}(\nmbObserversSubset)$ in Table~\ref{tab:validate_P_numerically} based on the real-world network behavior measured in Bitcoin (\cf~Section~\ref{sec:kit_data}).
We observe that $\PPV_{\nmbObservers}(\nmbObserversSubset)$ increases in \nmbObserversSubset for a fixed \nmbObservers. 
The proof for Theorem~\ref{th:predictive_precision} is given in Section~\ref{sec:proofs}.

\subsection{Instantiating \DistReception in Bitcoin} \label{sec:kit_data}

We now evaluate $\probGMO_\nmbObservers(\nmbObserversSubset)$ and $\PPV_{\nmbObservers}(\nmbObserversSubset)$ in Bitcoin for various combinations of the parameters $\nmbObserversSubset$, $\nmbObservers$, and $\timeDiff$.
We proceed as follows: first, we use the distribution of transaction propagation times derived from real-world measurements to instantiate the distribution \DistReception. 
By sampling probabilistic propagation times from this distribution, we then estimate $\probGMO_\nmbObservers(\nmbObserversSubset)$ and $\PPV_{\nmbObservers}(\nmbObserversSubset)$ for various numbers of observers \nmbObservers and transactions $\transaction_1$ and $\transaction_2$ sent various time differences \timeDiff apart.

\subheading{The KIT Dataset: } 
Neudecker \etal~\cite{neudecker-atc16} provide data from several measurements of the Bitcoin network and its transactions' and blocks' transmission. This data (referred to as the KIT dataset in the sequel) provides an empirical distribution of average transaction reception times within the Bitcoin network. In particular, it evaluates the PDF of the time it takes for a transaction to be forwarded through the network until it is received by an average node.  

The distribution of transaction reception times in the KIT dataset is depicted in \cref{fig:KIT_dist}. Here, the probability density function (PDF) of the average transaction's reception times is shown by the blue curve in \cref{fig:KIT_PDF}. We observe that most transactions are received within the first twenty seconds. Specifically, we observe that 90\% of all transactions are received after $20$ seconds, and 99\% of all transactions are received after at most $27.75$ seconds. Notice also the long tail of the distribution, indicating that some transactions take over 30 seconds to be received by certain nodes.

\subheading{Estimating the distribution of transactions' propagation time: } Based on the KIT dataset, we approximate the distribution of transactions' propagation time using the log-normal distribution  \linebreak
$\DistReceptionLN(\mu, \sigma^2)$ with mean value $\mu=1.973$ and variance $\sigma^2=0.585$. 
Our approximation allows us to evaluate the distribution more fine-granularly and not rely on data sampled at discrete time points. It also allows us to estimate transaction transmission times beyond the 30 seconds measured in the KIT dataset.

\subheading{Evaluating  $\probGMO_\nmbObservers$: }
We then use the log-normal distribution \linebreak $\DistReceptionLN(1.973, 0.585)$ to evaluate the probability $\probGMO_\nmbObservers$ for $\nmbObservers=5$ and $\nmbObservers=16$ observers. Additionally, we also evaluate the probability $\probGMO_\nmbObservers$ cumulatively for any subset of at least \nmbObserversSubset out of \nmbObservers observers, i.e., $\sum_{\nmbObserversSubset' \geq \nmbObserversSubset} \probGMO_\nmbObservers(\nmbObserversSubset')$.

\begin{table*}[tb]
    \begin{minipage}{.3\textwidth}
    \centering
    \footnotesize
    \scalebox{0.85}{\begin{tabular}{|c||r|r|r||r|r|}
        \hline
        \multirow{2}{*}{\timeDiff} & \multicolumn{3}{c||}{$\probGMO_{5}(\nmbObserversSubset)$} & \multicolumn{2}{c|}{$\sum_{\nmbObserversSubset'}\probGMO_{5}(\nmbObserversSubset')$} \\
        \cline{2-6}
        & $\nmbObserversSubset = 3$ & $\nmbObserversSubset =4$ & $\nmbObserversSubset = 5$ & $\nmbObserversSubset' \geq 3$ & $\nmbObserversSubset' \geq 4$ \\
        \hline
        \hline
        0 & 31.2\% & 15.6\% & 3.1\% & 50.0\% & 18.7\% \\
        \hline
        1 & 34.0\% & 21.5\% & 5.4\% & 60.9\% & 26.9\% \\
        \hline
        5 & 26.8\% & 39.3\% & 23.1\% & 89.2\% & 62.4\% \\
        \hline
        10 & 11.1\% & 37.2\% & 49.9\% & 98.2\% & 87.1\% \\
        \hline
        30 & 0.2\% & 7.2\% & 92.5\% & 100.0\% & 99.8\% \\
        \hline
        60 & 0.0\% & 0.8\% & 99.2\% & 100.0\% & 100.0\% \\
        \hline
        600 & 0.0\% & 0.0\% & 100.0\% & 100.0\% & 100.0\% \\
        \hline
    \end{tabular}}
    \caption{Estimate of $\probGMO_5(\nmbObserversSubset)$ based on the KIT distribution's approximation in Bitcoin.}
    \label{tab:validate_Q_numerically_smallN}
    \end{minipage}%
    \hspace{1em}
    \begin{minipage}{.3\textwidth}
    \centering
    \footnotesize
    \scalebox{0.85}{\begin{tabular}{|c||r|r|r||r|r|}
        \hline
        \multirow{2}{*}{\timeDiff} & \multicolumn{3}{c||}{$\probGMO_{16}(\nmbObserversSubset)$} & \multicolumn{2}{c|}{$\sum_{\nmbObserversSubset'}\probGMO_{16}(\nmbObserversSubset')$} \\
        \cline{2-6}
        & $\nmbObserversSubset = 8$ & $\nmbObserversSubset =12$ & $\nmbObserversSubset = 16$ & $\nmbObserversSubset' \geq 8$ & $\nmbObserversSubset' \geq 12$ \\
        \hline
        \hline
        0 & 19.6\% & 2.8\% & 0.0\% & 59.8\% & 3.8\% \\
        \hline
        1 & 17.6\% & 6.4\% & 0.0\% & 76.6\% & 9.7\% \\
        \hline
        5 & 2.1\% & 22.5\% & 0.9\% & 99.2\% & 61.5\% \\
        \hline
        10 & 0.0\% & 9.7\% & 10.8\% & 100.0\% & 95.3\% \\
        \hline
        30 & 0.0\% & 0.0\% & 78.0\% & 100.0\% & 100.0\% \\
        \hline
        60 & 0.0\% & 0.0\% & 97.4\% & 100.0\% & 100.0\% \\
        \hline
        600 & 0.0\% & 0.0\% & 100.0\% & 100.0\% & 100.0\% \\
        \hline
    \end{tabular}}
    \caption{Estimate of $\probGMO_{16}(\nmbObserversSubset)$ based on the KIT distribution in Bitcoin. }
    \label{tab:validate_Q_numerically_largeN}
    \end{minipage} 
    \hspace{1em}
    \begin{minipage}{.3\textwidth}
    \centering
    \footnotesize
    \scalebox{1.0}{\begin{tabular}{|c|c|@{}p{0.25cm}@{}|c|c|@{}p{0.25cm}@{}|c|c|}
        \cline{1-2} \cline{4-5} \cline{7-8}
        \multicolumn{2}{|c|}{$\nmbObservers=5$} & & \multicolumn{2}{c|}{$\nmbObservers=12$} & & \multicolumn{2}{c|}{$\nmbObservers=16$} \\
        \cline{1-2} \cline{4-5} \cline{7-8} 
        \nmbObserversSubset & $\PPV_{5}(\nmbObserversSubset)$ && \nmbObserversSubset & $\PPV_{12}(\nmbObserversSubset)$ && \nmbObserversSubset & $\PPV_{16}(\nmbObserversSubset)$ \\
        \cline{1-2} \cline{4-5} \cline{7-8}
        0 & 0.10\%  &&  6 & 50.98\%  &&  9 & 71.60\% \\
        1 & 4.59\%  &&  7 & 74.79\%  && 10 & 86.82\% \\
        2 & 29.73\% &&  8 & 90.63\%  && 11 & 95.27\% \\
        3 & 72.04\% &&  9 & 97.66\%  && 12 & 98.75\% \\
        \cdashline{7-8}    
        4 & 96.16\% && 10 & 99.66\%  && 14 & 99.98\% \\
        \cdashline{7-8}  
        5 & 99.90\% && 12 & 100.0\%  && 16 & 100.00\% \\
        \cline{1-2} \cline{4-5} \cline{7-8}
    \end{tabular}}
    \captionof{table}{valuating $\PPV_{\nmbObservers}(\nmbObserversSubset)$ using the KIT distribution in Bitcoin.} 
    \label{tab:validate_P_numerically}
    \end{minipage}%
\end{table*}

As shown in Table~\ref{tab:validate_Q_numerically_smallN}, when the time difference $\timeDiff$ between transactions is small (i.e., $\timeDiff \leq 5$ seconds), the likelihood that all $\nmbObserversSubset = 5$ observers receive $\transaction_1$ and $\transaction_2$ in the correct order is at most 23\%. In contrast, when the transactions are spaced by at least $\timeDiff \geq 60$ seconds, this probability exceeds 99\%, indicating a strong temporal separation is necessary for consistent ordering across all observers. This intuitively confirms the folklore belief that longer time differences between the sending times of two transactions yield more robustness against accidental re-orderings.  
Table~\ref{tab:validate_Q_numerically_largeN} shows function $\probGMO_\nmbObservers$ evaluated for $\nmbObservers=16$ observers. Here, we observe that two transactions sent less than $\timeDiff\leq 30$ seconds apart are received by all $16$ observers with probability less than 78\%. In other words, 22\% of the transactions with $\timeDiff \leq 30$ are received by our observers in an inconsistent order.  
Nevertheless, we observe that receiving two transactions with $\timeDiff\geq30$ by four out of five (\cf~Table~\ref{tab:validate_Q_numerically_smallN}) and by 12 out of 16 observers (\cf~Table~\ref{tab:validate_Q_numerically_largeN}), i.e., a vast majority of nodes, in order is (nearly) guaranteed.

Overall, we observe that a larger number of observers \nmbObservers reduces the probability that they all witness one certain order, i.e., 
\begin{equation}
    \nmbObservers_1 > \nmbObservers_2 \ \Rightarrow \ \probGMO_{\nmbObservers_1}(\nmbObservers_1) \leq \probGMO_{\nmbObservers_2}(\nmbObservers_2). 
\end{equation}

However, a larger number of observers witnessing the same order (i.e., $m$) also increases the precision (or reliability) when detecting order inconsistencies (cf.~Table~\ref{tab:validate_P_numerically}). 

Naturally, increasing the number of observers \nmbObservers yields more robust results. That is, having more observers decreases the probability that a displacement attack remains unnoticed. However, a high number of observers also reduces the probability that all nodes reach agreement on their mempools' order. As we model the reception of a transaction as a probabilistic process, i.e., sampling the propagation times from a probability distribution $\timeVar_{\observer}(\transaction) \sim \DistReception$, the probability that all observers unanimously report a distinct order reduces as \nmbObservers increases.

\begin{figure}
    \centering
    \includegraphics[width=0.47\linewidth]{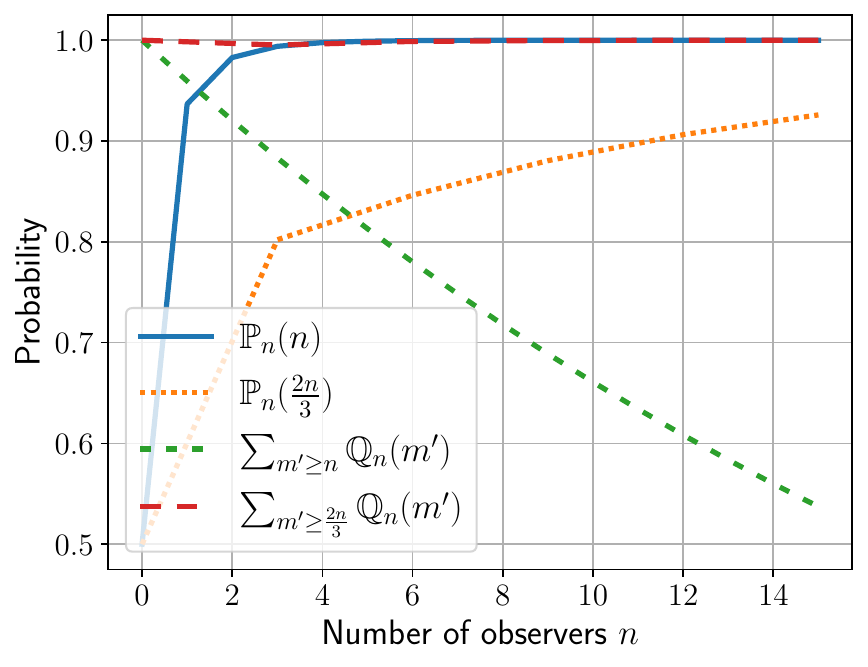}
    \caption{$\sum_{\nmbObserversSubset'\geq \nmbObserversSubset} \probGMO_{\nmbObservers}(\nmbObserversSubset')$ / $\PPV_{\nmbObservers}(\nmbObserversSubset)$ w.r.t. the number of observers \nmbObservers. }
    \label{fig:precision_vs_recall}
\end{figure}

Finally, we evaluate both $\sum_{\nmbObserversSubset'\geq \nmbObserversSubset}\probGMO_{\nmbObservers}(\nmbObserversSubset')$ and $\PPV_{\nmbObservers}(\nmbObserversSubset)$, which can be used to measure the recall and the precision of mempool auditing schemes, as a function of the number of observers \nmbObservers in the network in Figure~\ref{fig:precision_vs_recall}. On the one hand, we see that increasing the number of observers drastically decreases the probability that all observers witness a pair of two transactions $\transaction_1, \transaction_2$ in the same order (i.e., $\sum_{\nmbObserversSubset'\geq \nmbObserversSubset}\probGMO_{\nmbObservers}(\nmbObserversSubset')$). On the other hand, increasing \nmbObservers yields a more precise prediction of the original sending order of $\transaction_1$ and $\transaction_2$ (i.e., $\PPV_{\nmbObservers}(\nmbObserversSubset)$). Relying on a $\nmbObserversSubset=\frac{2\nmbObservers}{3}$ majority improves $\sum_{\nmbObserversSubset'\geq \nmbObserversSubset}\probGMO_{\nmbObservers}(\nmbObserversSubset')$ but seriously deteriorates probability $\PPV_{\nmbObservers}(\nmbObserversSubset)$ of the scheme.
Additionally, Figure~\ref{fig:precision_vs_recall} shows that the ideal number of observers for precise and reliable observations is rather small, e.g., two to six. This contrasts with the large number of observers ($n=24$) used in  Mempool.space~\cite{mempool_space}.

\subsection{Displacement across Blocks}\label{sec:DaB}

We now determine the probability that a transaction gets displaced across blocks. 
Typically, a displacement to a different block than expected is triggered when a single transaction is not only inverted in order with another transaction but with a sequence of transactions.
Here, we refer to a block \block as an ordered sequence of transactions, i.e., $\block^{(i)}=[\transaction^{(i)}_1, \transaction^{(i)}_2, \dots\,]$. Displacement across blocks denotes the event that a transaction \transaction should be scheduled according to the deterministic algorithm (\cf~\cref{sec:problem_statement}) in block $\block^{(i)}$ but is incorporated in another block $\block^{(j)}$ with $i < j$ appearing after $\block^{(i)}$. 

Instances of displacement across blocks are typically detectable by comparing a proposed block with its expected view. Specifically, each observer receives new transactions and stores them in its local mempool, i.e., it maintains a list of unprocessed transactions. Based on its local mempool, every observer has an expected view on the next block, i.e., the expected set of transactions and their order of execution. Then, once a new block $\block^{(i)}$ for slot $(i)$ is mined, we can compare $\block^{(i)}$ with the expected blocks $\block^{(i)}_\observer$ of the observers. 

\begin{proposition} \label{prop:DAB}
    Given \nmbObservers observers that each have an expected view $\block^{(i)}_\observer$ on a certain block and the actual block $\block^{(i)}$, a transaction \transaction has been displaced across blocks if 
    \begin{equation}
        \forall \observer \in \SObservers: \transaction \in \block^{(i)}_\observer \ \land \ \transaction \in \block^{(j)},\ j>i. 
    \end{equation}
\end{proposition}

In other words, if a transaction \transaction that is apparent in the expected blocks $\block_\observer$ of all observers but is missing in \block, then \transaction appears to have been displaced across blocks. 
However, as shown in Section~\ref{sec:network_behavior}, two honest nodes might receive transactions in a different order. That is, not every transaction that is expected in $\block_\observer$ and missing in \block is an intentional displacement. In fact, the probability that a transaction \transaction is omitted by an honest miner without any intention to displace \transaction can be determined as shown in \cref{th:DAB}.  

\begin{theorem}[] \label{th:DAB}
    Given a transaction $\transaction$ and \nmbObservers observers, the probability that a sequence of transactions $(\transaction_i)_{i}$, which have all been sent at least \timeDiff time after \transaction, i.e., $\sendTime{\transaction_i} \geq \sendTime{\transaction} + \timeDiff,\ \forall i$, are all received before $\transaction$ is:
    \begin{equation}
        \pr{\bigwedge_{\observer \in \SObservers}\left( \bigwedge_{i} \receivedBeforeBy{\transaction_i}{\transaction}{\observer} \right)} 
        = \Big(1 - \CDF{\timeVar_{\observer}(\transaction)}{\censorTime}\Big)^{n}   
    \end{equation}
\end{theorem}

Recently received transactions are excluded from this definition; a transaction must be received at least \censorTime time ago before it is considered to be displaced (see full proof in Section~\ref{sec:proofs}).

\section{Empirical Validation} \label{sec:empirical_validation_network}

In this section, we empirically evaluate our analysis presented in Section~\ref{sec:network_model} using two popular blockchains, Bitcoin and Ethereum. 
Specifically, we evaluate the probability that at least \nmbObserversSubset out of \nmbObservers blockchain nodes receive $\transaction_1$ before $\transaction_2$, i.e., $\sum_{\nmbObserversSubset'\geq \nmbObserversSubset} \probGMO_\nmbObservers(\nmbObserversSubset)$. 

For this purpose, we set up 16 Bitcoin and Ethereum nodes in different regions around the globe (cf.~Table~\ref{tab:server_location} in Appendix~\ref{ap:tables}) to record transactions' reception times and their order of arrival. Our nodes were hosted by Hetzner\footnote{We initially used AWS but faced stability issues with nodes crashing, so we used Hetzner for more reliable measurements.} cloud~\cite{hetzner_cloud}; each server featured four dedicated vCPU kernels, 16GB RAM, SSD storage, an AMD EPYC-Milan processor, and was reachable by an IPv4 address. All measurements were conducted on servers running Ubuntu 24.04 LTS. We used \emph{bitcoind 25.0} as our Bitcoin client and the \emph{geth 1.15.11} client for Ethereum. Throughout our measurements, each observer logged all incoming transactions and their exact reception time between 30.06. and 08.07.2025, by a total of 16 geographically dispersed observers (see Table~\ref{tab:server_location_EUS} in Appendix~\ref{ap:tables}).

\subsection{Real-world Evaluation in Bitcoin} \label{sec:bc_eval}

In Bitcoin, our $\nmbObservers=16$ observers recorded in total \numprint{3134824} transactions over the course of 196 hours. Before evaluating our findings, we first report the detailed reception statistics in Table~\ref{tab:res_reception_general_both_LS}. 
While one would expect all transactions to be eventually received by any node in the network, we observe that some transactions were only partially received by a subset of our 16 observers. 
Here, we observe that out of all transactions, only 93.93\% were received by every observer. Further, 99.08\% of all transactions were received by a subset of at least $\nmbObserversSubset \geq 12$ nodes, and 99.34\% were received by a majority of at least $\nmbObserversSubset \geq 9$ nodes. 
On average, each node received $98.9\% \pm 0.15\%$ of the recorded transactions. 
We believe that the fraction of transactions received only by a subset of our $16$ observers is due to transactions that have already circulated through the network for a while and are not forwarded again by all nodes, or due to transactions that have not reached all 16 nodes yet. Based on the recorded transactions, we compute the pairwise Mutual Information (MI) score between observers' reception times, and find an average MI of 0.15 nats, indicating that reception times are practically independent, confirming our assumption in Section~\ref{sec:network_behavior}.  

We now use our recorded data to empirically validate our analysis. First, we determine an empirical estimate of $\probGMO_\nmbObservers(\nmbObserversSubset)$. That is, we are interested in the fraction of transaction pairs for which our observers agree on their order of reception. For that purpose, we sample \numprint{100000} random pairs of transactions $\bigl(\transaction^{(i)}_0, \transaction^{(i)}_1\bigr)$ with $i \in I:= [1, \numprint{100000}]$, i.e., \numprint{200000} transactions that are sampled uniform at random from our recorded transaction sets and randomly paired. 
For each pair $i$, we then determine the number of observers that agree on an order, i.e., 
\begin{align}
    &\mathcal{C}_{order}(\nmbObserversSubset) = \frac{1}{|I|} \cdot \Bigl| \Bigl\{ i \in I \ \big| \ \nmbObserversSubset \geq X(i) \Bigr\}\Bigr| \label{eq:C_order}
\end{align}
with $X(i)$ denoting the number of observers that witnessed the most frequent order of reception: 
\begin{align}
        &X(i) = \max_{j\in\{0,1\}}\bigl|\bigl\{ \observer \in \SObservers \ | \ \receivedBeforeBy{\transaction^{(i)}_j}{\transaction^{(i)}_{1-j}}{\observer} \bigr\}\bigr|.
\end{align}
In other words, $\mathcal{C}_{order}(\nmbObserversSubset)$ is the fraction of transaction pairs that satisfy the condition $\nmbObserversSubset \geq X(i)$, i.e., the fraction of pairs that have been received in the same order by at least \nmbObserversSubset out of \nmbObservers observers. Hence, $\mathcal{C}_{order}(\nmbObserversSubset)$ is the empirical equivalent to our probability $\sum_{\nmbObserversSubset'\geq \nmbObserversSubset} \probGMO_\nmbObservers(\nmbObserversSubset')$\footnote{Since the ground truth for the sending time $\timeVar$ of transactions is unavailable, we cannot determine the exact sending time of individual transactions. However, because sending and reception times follow similar distributions, discrepancies in propagation times are expected to average out across 100,000 pairs.}.

\cref{tab:res_reception_general_both_LS} shows the evaluation results of $\mathcal{C}_{order}(\nmbObserversSubset)$ with respect to $\nmbObserversSubset$ in Bitcoin. Here, we sampled the \numprint{100,000} transaction pairs from (1) the complete dataset (\emph{All Txs}) and (2) the subset of transactions that were fully received by all 16 observers (\emph{\WRTxs}). Our results show that 88.23\% of the randomly drawn transaction pairs from \emph{All Txs} are witnessed in the same order by all 16 observers. By restricting our dataset to transactions from \emph{\WRTxs}, we observe that 99.98\% of the sampled transaction pairs were received in order.

\begin{figure*}[!htb]
    \begin{minipage}{.29\linewidth}
    \centering
    \includegraphics[width=\linewidth]{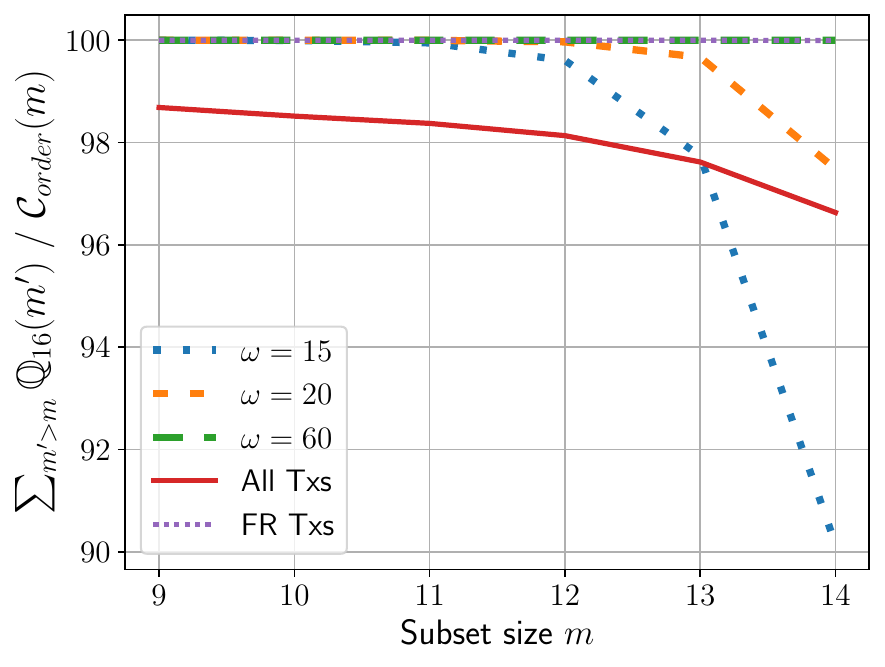}
    \caption{Comparison of $\sum_{\nmbObserversSubset'\geq \nmbObserversSubset} \probGMO_{16}(\nmbObserversSubset')$ (dashed lines) with the empirically measured $\mathcal{C}_{order}(\nmbObserversSubset)$ (solid/dotted lines) using the complete dataset (\emph{All Txs}) and the set of transactions received by all observers (\emph{\WRTxs}) measured by 16 Bitcoin observers.}
    \label{fig:QvsC_BC}
    \end{minipage}%
    \hspace{1em}
    \begin{minipage}{.29\linewidth}
    \centering
    \includegraphics[width=\linewidth]{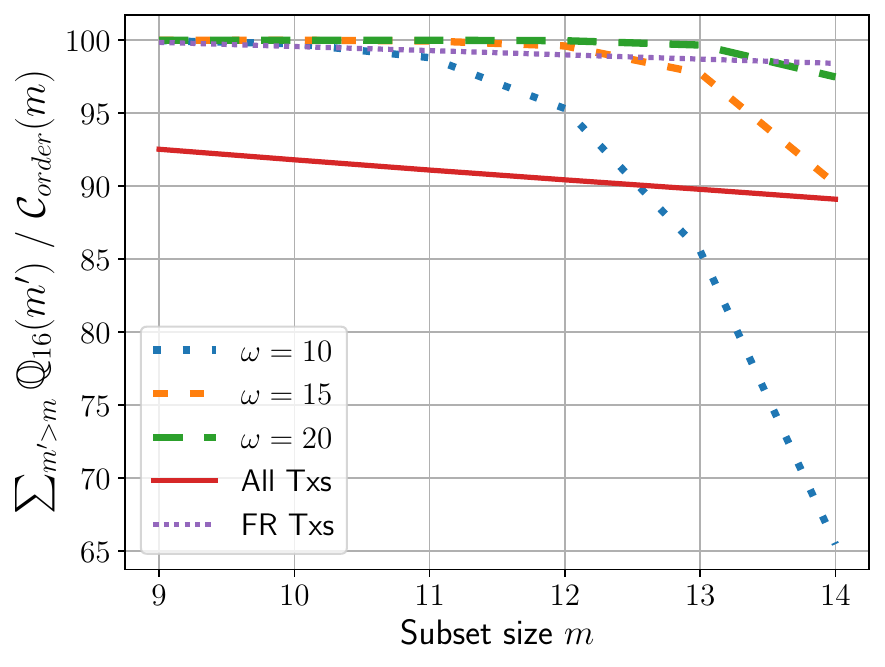}
    \caption{Comparing  $\sum_{\nmbObserversSubset'\geq \nmbObserversSubset} \probGMO_{20}(\nmbObserversSubset')$ (dashed lines) with the empirical measurement  $\mathcal{C}_{order}(\nmbObserversSubset)$ (solid/dotted lines) using the full dataset (\emph{All Txs}) and the transactions received by all observers (\emph{\WRTxs}) measured by 16 Ethereum observers.}
    \label{fig:QvsC_ETH}
    \end{minipage} 
    \hspace{1em}
    \begin{minipage}{.3\linewidth}
    \includegraphics[width=\linewidth]{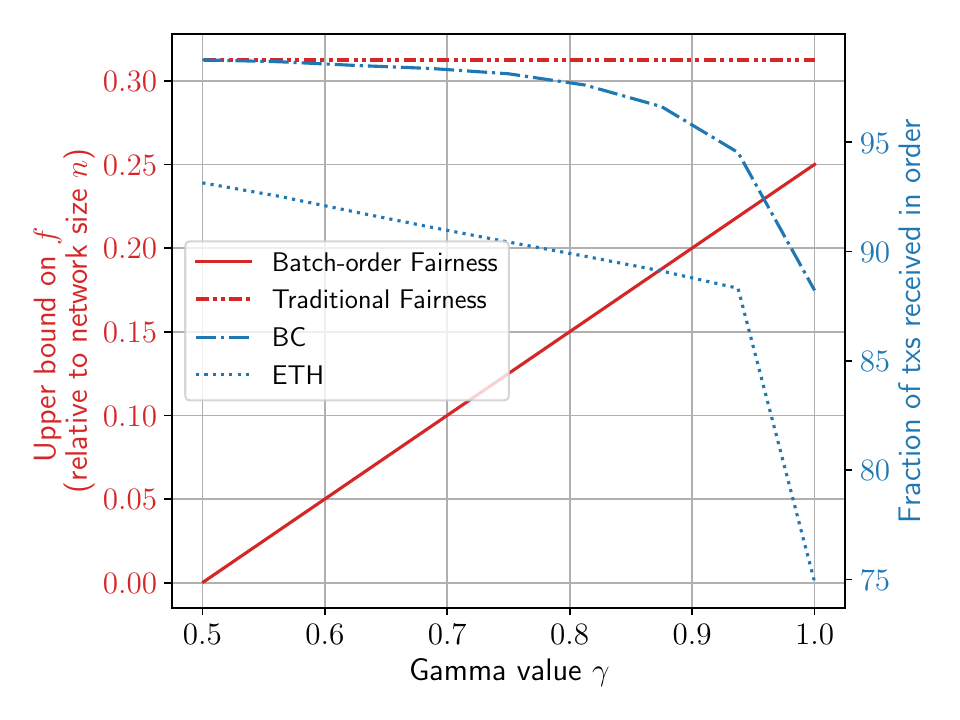}
    \captionof{figure}{Comparison of batch-order fairness schemes (e.g., Themis\cite{DBLP:conf/ccs/KelkarDLJK23}) vs. traditional fair-ordering schemes (e.g., Pompe~\cite{DBLP:conf/osdi/ZhangSCZA20} and Wendy~\cite{DBLP:conf/aft/Kursawe20}). We see that Themis only offers strong guarantees to a fraction of the transactions.}
    \label{fig:themis_bounds}
    \end{minipage}%
\end{figure*}

We then compare our empirical findings in the Bitcoin blockchain with the probabilities from \cref{tab:validate_Q_numerically_largeN} (\cf~Section~\ref{sec:kit_data}) in Figure~\ref{fig:QvsC_BC}. Here, the dashed lines denote the cumulative probability that at least \nmbObserversSubset out of \nmbObservers observers receive two transactions sent \timeDiff seconds apart in the rightful order, i.e., $\sum_{\nmbObserversSubset'\geq \nmbObserversSubset} \probGMO_\nmbObservers(\nmbObserversSubset')$ (\cf~\cref{sec:network_behavior}). Further, the solid, red, and lilac lines denote our measurement results as shown in \cref{tab:res_reception_general_both_LS}, i.e., $\mathcal{C}_{order}(\nmbObserversSubset)$. 
The dashed lines show that the probability that $\nmbObserversSubset\geq14$ observers witness the same order is approximately 97.5\% when $\timeDiff=20$ while our empirical results reveal that transaction pairs in practice are received in the same order by $\nmbObserversSubset\geq14$ observers with a comparable probability of 96.63\% shown by the red plot (\emph{All Txs}). Further, the violet plot (\emph{\WRTxs}) reveals the probability of fully-received transactions to be almost guaranteed. This conforms with our analytical results for $\timeDiff \geq 60$. 

Note that our empirical results on the Bitcoin network are based on measurements conducted over the course of 196 hours. Hence, on average, our randomly sampled transaction pairs used to determine $\mathcal{C}_{order}(\nmbObserversSubset)$ have a time difference of $\timeDiff \approx \numprint{236028}$ seconds. 

In Appendix~\ref{ap:EUS_results}, we additionally present selected results using observers located exclusively in the US and Europe, reflecting the current deployment landscape in which the majority of Bitcoin and Ethereum observers\footnote{According to data from \href{https://bitnodes.io/}{bitnodes.io}, \href{https://www.ethernodes.org/countries}{ethernodes.org}, and \href{https://mempool.space/lightning/group/the-mempool-open-source-project}{mempool.space}.}---as well as most Mempool.space~\cite{mempool_space} observers---are situated in these regions. Our findings remain consistent regardless of whether observers are restricted to the US/EU or distributed more globally, including regions such as Asia. Furthermore, our results are robust across different measurement periods.\footnote{In Appendix~\ref{ap:first_measurement_results}, we include data from an initial measurement campaign conducted in October 2024 using 20 Bitcoin nodes, which are also consistent with our findings.}

\begin{table}[tbp]
\centering
\scalebox{0.82}{\begin{tabular}{|c||r|r|r||r|r|r|}
        \cline{1-7} 
        & \multicolumn{3}{c||}{Bitcoin (\numprint{3134824} Txs)} & \multicolumn{3}{c|}{Ethereum (\numprint{6552228} Txs)} \\
        \cline{1-7}
        \multirow{2}{*}{\nmbObserversSubset} & \multicolumn{1}{c|}{Received} & \multicolumn{2}{c||}{$\mathcal{C}_{order}(\nmbObserversSubset)$} & \multicolumn{1}{c|}{Received} & \multicolumn{2}{c|}{$\mathcal{C}_{order}(\nmbObserversSubset)$} \\
         & \multicolumn{1}{c|}{Txs} & \multicolumn{1}{c|}{All Txs} & \multicolumn{1}{c||}{\WRTxs} & \multicolumn{1}{c|}{Txs} & \multicolumn{1}{c|}{All Txs} & \multicolumn{1}{c|}{\WRTxs} \\
        \hline
        \hline
        $\geq8$ nodes & 99.38\% & 98.77\% & 100.0\%  & 96.49\% & 93.21\% & 100.00\%  \\
        $\geq9$ nodes & 99.34\% & 98.68\% & 100.0\%  & 96.29\% & 92.57\% & 99.86\%  \\
        \hdashline
        $\geq10$ nodes & 99.26\% & 98.52\% & 100.0\% & 96.08\% & 91.86\% & 99.57\%  \\
        $\geq11$ nodes & 99.19\% & 98.37\% & 100.0\% & 95.87\% & 91.15\% & 99.29\%  \\
        \hdashline
        $\geq12$ nodes & 99.08\% & 98.13\% & 100.0\% & 95.26\% & 90.47\% & 99.00\%  \\
        $\geq13$ nodes & 98.83\% & 97.62\% & 100.0\% & 95.63\% & 89.82\% & 98.71\%  \\
        \hdashline
        $\geq14$ nodes & 98.34\% & 96.63\% & 100.0\% & 95.34\% & 89.14\% & 98.42\%  \\
        $\geq15$ nodes & 97.28\% & 94.53\% & 99.99\% & 94.96\% & 88.33\% & 98.13\%  \\
        \hdashline
        16 nodes       & 93.93\% & 88.23\% & 99.98\% & 87.49\% & 74.92\% & 97.84\%  \\
        \hline
    \end{tabular}}
    \caption{
        \emph{Received Txs} denotes the percentage of transactions recorded by at least \nmbObserversSubset out of $\nmbObservers=16$ observers in Bitcoin and Ethereum. \emph{$\mathcal{C}_{order}(\nmbObserversSubset)$} denotes the fraction of transactions received in a consistent order by $\nmbObserversSubset$ nodes. Results labeled with \emph{\WRTxs} are based on the intersection of transactions received by all 16 observers. Results labeled with \emph{All Txs} include all transactions.
    }
    \label{tab:res_reception_general_both_LS}
\end{table}

\subsection{Real-world Evaluation in Ethereum} \label{sec:eth_eval}

Our evaluation results in the Ethereum blockchain are shown in Table~\ref{tab:res_reception_general_both_LS} and encompass a period of 196 hours, during which we recorded a total of \numprint{6552228} Ethereum transactions. Out of all analyzed transactions, nearly 87.5\% were received by all 16 nodes, 95.26\% were received by a subset of 12 nodes, and 96.29\% by a majority of at least nine nodes. Each node received, on average, $96.32\% \pm 0.83\%$ of the recorded Ethereum transactions. However, one of the 16 observers received only 89.67\% of the transactions. 

Based on this analysis, we measure $\mathcal{C}_{order}(\nmbObserversSubset)$ (\cf~\cref{eq:C_order}), i.e., the empirical estimate of $\sum_{\nmbObserversSubset'\geq \nmbObserversSubset} \probGMO_\nmbObservers(\nmbObserversSubset')$, in Ethereum. Our results are shown in Table~\ref{tab:res_reception_general_both_LS}. Here, we observe that the fraction of transaction pairs sampled from \emph{All Txs} that were witnessed in the same order by all observers is only 74.92\%, and the fraction witnessed by a subset of at least 12 out of 16 observers is only  90.47\%. 
Recall that our Ethereum nodes did not receive all recorded transactions; needless to mention, nodes cannot take a stance on the pairwise ordering of two transactions if one or both were not received. If we sample transaction pairs from \emph{\WRTxs}, we observe that $97.84\%$ of these transactions were received in the same order by all 16 observers, and roughly $\geq99.0\%$ by at least 12 observers.

Note that there exists no ground truth for the distribution of our transactions' propagation times, which prevents us from directly comparing our empirical results on $\mathcal{C}_{order}(\nmbObserversSubset)$ with its analytical counterpart. However, Figure~\ref{fig:QvsC_ETH} shows a best-effort approximation on the comparison between $\mathcal{C}_{order}(\nmbObserversSubset)$ (solid plots) and $\sum_{\nmbObserversSubset'\geq \nmbObserversSubset} \probGMO_\nmbObservers(\nmbObserversSubset')$ (dashed plots) instantiated with the distribution of propagation times $\DistReception=\DistReceptionLN(1.973, 0.585)$ (\cf~\cref{sec:kit_data}). 
Our empirical results for $\mathcal{C}_{order}(\nmbObserversSubset)$ using \emph{All Txs} resemble those for $\sum_{\nmbObserversSubset' \geq \nmbObserversSubset} \probGMO_\nmbObservers(\nmbObserversSubset')$ with $\timeDiff \in [10,15]$ seconds. Similarly, we observe that our results on $\mathcal{C}_{order}(\nmbObserversSubset)$ using only transactions from \emph{\WRTxs} closely follow our empirical results recorded in Bitcoin and conform to $\sum_{\nmbObserversSubset'\geq \nmbObserversSubset} \probGMO_\nmbObservers(\nmbObserversSubset')$ with $\timeDiff\geq 20$. 

Specifically, our results show that transactions received by all observers are received in a consistent order among our observers with high probability. This confirms our analytical results for large time differences $\timeDiff \gg 600$ shown in Table~\ref{tab:validate_Q_numerically_smallN} and Table~\ref{tab:validate_Q_numerically_largeN}, respectively. 

Overall, our results suggest a more consistent reception of transactions in the Bitcoin network compared to the Ethereum network. 
That is, 99.34\% of the Bitcoin transactions and 96.29\% of the Ethereum transactions were received by a majority, i.e., $\nmbObserversSubset\geq9$, of observers. 
In general, we suspect the unreliable delivery of transactions is caused by: (1) the connectivity of our observers in the respective blockchain's network, (2) the limited duration of our measurement, or (3) unstable network connections of our cloud nodes. 
Furthermore, we suspect the specific differences between Bitcoin and Ethereum to be caused by messages sent in off-chain networks, such as flash loan transactions, and/or a large number of second-layer applications in Ethereum  (\cf~Appendix~\ref{sec:limitations}). 

\section{Interpretation of Findings} \label{sec:interpretation_of_findings}

\noindent \textbf{How likely are independent observers to witness the same set of transactions? (RQ1).}
In Bitcoin, $93.93\%$ of the recorded transactions were observed by all 16 observers, and $99.08\%$ were observed by at least 12 out of 16. In contrast, in Ethereum, only $87.49\%$ of the recorded transactions were received by all observers, and $95.26\%$ by at least 12 out of 16. 

\subheading{Can Mempool Auditing dismiss displacement attacks within blocks? (RQ2).}
Our results indicate that detecting transaction displacements primarily depends on the sending time difference $\timeDiff$ and the number of observers $\nmbObservers$. Larger $\timeDiff$ values increase detection certainty. 
Using the KIT distribution of Bitcoin propagation times $\DistReception$, Tables~\ref{tab:validate_Q_numerically_smallN} and \ref{tab:validate_Q_numerically_largeN} show that for $\timeDiff \geq 30$, most observers (4/5 and 12/16, respectively) agree on transaction order. Thus, if two transactions are sent within 30 seconds,  displacement attacks can be reliably detected. Empirical measurements confirm this: while some transactions are inconsistently received, pairs observed by all 16 observers almost always preserve order. In contrast, displacements with short gaps ($\timeDiff \leq 30$s) cannot always be detected.

\subheading{Can Mempool Auditing help dismiss displacement across blocks? (RQ3).}
Our third research question examines when transaction displacement across blocks can be confidently detected. Assuming all honest, rational nodes (i) order mempool transactions by arrival time and (ii) apply a public deterministic scheduling algorithm, one can estimate execution order by monitoring arrivals and comparing the block’s contents to a node’s expected set.

However, our results (\cref{tab:res_reception_general_both_LS}) show limited certainty: 0.92\% of Bitcoin transactions and 4.74\% of Ethereum transactions were missed by 4/16 nodes. Thus, cross-block displacement cannot generally be detected with certainty (RQ3). 

Censorship is implicitly covered in this analysis, since omitting a transaction when minting a block is equivalent to displacing it across blocks. 
Note that protocol parameters like block rate do not change this conclusion but affect interpretation: cross-block displacement is harder to detect in high block-rate chains (e.g., Ethereum) than in low block-rate ones (e.g., Bitcoin) due to the shorter block interval.

\subheading{How do batch-order fair-ordering schemes perform compared to traditional fair-ordering approaches? (RQ4)}
Kelkar \etal~\cite{DBLP:conf/crypto/Kelkar0GJ20} introduced \emph{batch-order fairness}, which ensures fair ordering if a fraction $\gamma$ of honest nodes ($\tfrac{1}{2} < \gamma \leq 1$) receive two transactions in the same order. This guarantee is stronger than traditional fair-ordering properties (e.g., Pompe~\cite{DBLP:conf/osdi/ZhangSCZA20}, Wendy~\cite{DBLP:conf/aft/Kursawe20}), which allow a transaction $t_2$ to be output even if $t_1$, observed earlier by all, is withheld—thus enabling censorship.
Themis~\cite{DBLP:conf/ccs/KelkarDLJK23}, a prominent fair-ordering scheme, achieves $\gamma$-batch-order fairness but with reduced fault tolerance: it tolerates $f < n \tfrac{2\gamma-1}{2(\gamma+1)}$ corrupted nodes, compared to the traditional $f < \tfrac{n}{3}$.
Our analysis provides the first empirical assessment of $\gamma$, relating the adversarial bound $f(\gamma)$ to the observed fraction of nodes that agree on order ($\gamma = m/\nmbObservers$). Figure~\ref{fig:themis_bounds} shows that the strongest guarantee ($\gamma=1$, tolerating $f < \nmbObservers/4$) applies to 88.23\% of Bitcoin and 74.92\% of Ethereum transactions. A weaker guarantee ($\gamma=0.75$, 12/16 observers agreeing) covers 98.13\% of Bitcoin and 90.47\% of Ethereum transactions, but tolerates only $f < 0.143\nmbObservers$.
Overall, Figure~\ref{fig:themis_bounds} demonstrates that while batch-order fair-ordering schemes can provide interesting fairness guarantees, their applicability is limited to a subset of transactions.

\section{Threats to Validity} \label{sec:limitations}

\noindent \textbf{Non-uniform Distribution of Observers: } Our analysis assumes that observers are randomly distributed and not directly connected, ensuring that their information reception times are independent (Section~\ref{sec:network_behavior}). If this assumption does not hold, the reception times recorded by different observers for the same transactions may become correlated, which would further reduce the reliability of the auditing process, as the observed views would be more biased.

\subheading{Off-chain Network Events and Flash Loans.} \label{sec:limitation_offchain_txs}
A limitation of mempool auditing is the inability to detect transactions that are exchanged off-chain, such as flash loan transactions. 
First, since observers of the blockchain network do not receive off-chain transactions, they cannot log their reception times. 
Second, as off-chain transactions are (exclusively) queued up by a dedicated flash loan server and maintained in the server's mempool, assembling a block with either on-chain or off-chain transactions will naturally be recognized as an instance of transactions' displacements by other observers. 
While it is challenging to obtain reliable estimates of transaction counts and/or propagation characteristics from existing second-layer (L2) solutions, we measured the fraction of blobs, i.e., transaction types introduced in EIP-4844~\cite{eip_4844, eip_4844_git} and predominantly used to aggregate/pack L2 commitments onto the L1 blockchain, among all transactions in \numprint{500} randomly selected Ethereum blocks generated on 25.11.2025. 
Our measurements show that L2 blobs account for up to 6\% of L1 transactions. This estimate, together with Ethereum's higher throughput of $2.1\times$ that of Bitcoin, could explain the slightly lower consistency observed in Ethereum (Section~\ref{sec:eth_eval}).
  
Notice, however, that mempool auditing can be easily extended to monitor off-chain transactions as well to address this limitation. This can be achieved, e.g., by accepting connections from users that operate off-chain to our on-chain observers. Such observers would then offer an API to, e.g., users with flash loan transactions. Once a user intends to issue an off-chain transaction, i.e., a transaction that is sent to a flash loan server, the user additionally forwards it to a set of (off-chain) observers.

\subheading{Network \& Block Stuffing Attacks. }
A third limitation of mempool auditing arises from an adversary's ability to interfere with the propagation of honest transactions in the network. A network adversary that delays transaction dissemination---e.g., by dropping incoming transactions or by flooding the network with a large volume of transactions---may distort transaction reception times and potentially manipulate the observed ordering for its own benefit.

A concrete example is the block-stuffing attack~\cite{block_stuffing_attack}, in which the adversary injects a burst of high-fee transactions that saturate a block's gas limit, preventing the inclusion of other transactions. By flooding the network with such transactions, the adversary can delay the execution of honest users' transactions. If an observer $O$ is targeted by such an attack, real-time auditing---where decisions must be made within fixed time windows---may produce delayed or inconsistent mempool snapshots relative to those observed by other observers. This discrepancy can lead to cross-block displacement. Block-stuffing attacks can be mitigated through fee-rate ``smoothing''~\cite{DBLP:conf/atal/DamleSG24,bitcoin_cpfp,DBLP:journals/corr/abs-1901-06830}. For example, Child Pays For Parent~\cite{bitcoin_cpfp} enables boosting low-fee transactions by attaching a higher-fee child transaction that spends their outputs. Such mechanisms can be incorporated in the deterministic algorithm (\cf~Det.~Alg.,  Section~\ref{sec:problem_statement}) to enforce consistent behavior across miners.

Addressing the broader class of network-level attacks typically requires robust communication protocols that withstand adversarial interference~\cite{DBLP:conf/sp/AlbrechtAAKMW24}. Possible measures include increasing node fanout in graphs to ensure better connectivity among nodes, etc.

\section{Conclusion}

In this paper, we present the first in-depth analysis of the effectiveness of mempool-auditing in detecting transaction displacement attacks by malicious miners. Our analysis specifically examines the guarantees that mempool-based detection offers in realistic scenarios, such as those in the current Bitcoin and Ethereum ecosystems. Our findings indicate that while mempool auditing can assist in identifying transaction displacement attacks by miners, relying solely on this method to assess miners’ reputations—as some in the community advocate—does not provide definitive proof of malicious actions. In fact, our results suggest that mempool auditing may lead to false accusations against miners, with a probability exceeding 40\% in some settings. 

\section*{Acknowledgements}
This work is funded by the Deutsche Forschungsgemeinschaft (DFG, German Research Foundation) under Germany’s Excellence Strategy - EXC 2092 CASA - 390781972.

\bibliographystyle{ACM-Reference-Format}  
\bibliography{./bib.bib}       

\appendix
\section{Additional Related Work}  \label{sec:apdx_rel_work}

The literature features a number of contributions that analyze transaction displacement attacks and propose various mitigation approaches. An overview of common displacement attacks and defenses is given by Eskandari \etal~\cite{DBLP:conf/fc/EskandariMC19}, Baum \etal~\cite{DBLP:conf/fc/BaumCDFG22}, and Zhou \etal~\cite{DBLP:conf/sp/ZhouXECWWQWSG23}. In particular, Eskandari \etal~\cite{DBLP:conf/fc/EskandariMC19} categorize existing attack vectors into three classes \textit{displacement}, \textit{insertion}, and \textit{suppression}, all of which fall within our broader definition of displacement attacks. Existing solutions to mitigate the threat typically either (i) prevent the leakage of sensitive data until the corresponding transaction is executed or (ii) establish an order of execution that is considered fair. 
The first type of defenses \emph{commit \& reveal} is a commitment scheme and belongs to the former class. 
It attempts to thwart displacement and censoring attacks by blinding the transaction's data, i.e., its value and/or receiver. Then, after the transaction's execution is scheduled and cannot be modified, the hidden data is revealed. Libsubmarine~\cite{libsubmarine} is an open-source smart contract library that belongs to the class commit \& reveal schemes. It leverages a technique called Submarine Commitments proposed and first analyzed by Breidenbach \etal~\cite{DBLP:conf/uss/BreindenbachDTJ18}. 
Another line of work by Kursawe~\cite{DBLP:conf/aft/Kursawe20} and Kelkar \etal~\cite{DBLP:conf/crypto/Kelkar0GJ20,DBLP:conf/asiapkc/KelkarDK22,DBLP:conf/ccs/KelkarDLJK23} attempts to thwart front-running by establishing a fixed ordering of transactions' execution. Both Kursawe and Kelkar \etal~propose to establish a \emph{fair ordering} of transactions. In particular, Kelkar \etal~\cite{DBLP:conf/crypto/Kelkar0GJ20} proposed a consensus property coined \emph{transaction order-fairness} in the permissioned setting.  In subsequent work, Kelkar \etal~\cite{DBLP:conf/asiapkc/KelkarDK22} extended this property to the permissionless setting. 
Finally, Kelkar \etal~\cite{DBLP:conf/ccs/KelkarDLJK23} propose Themis---an integrative scheme to accomplish a fair ordering of transactions in permissioned blockchains based on the notion of batch-order fairness~\cite{DBLP:conf/crypto/Kelkar0GJ20} and reliant on a consistent ordering of transactions. 

Unfortunately, these existing solutions are impractical for existing blockchains. Namely, they either cannot support real-time detection/prevention or are not generically applicable to different blockchains/applications. 
For instance, commit \& reveal schemes such as submarine commitments~\cite{DBLP:conf/uss/BreindenbachDTJ18, libsubmarine} cannot function in real-time. Since the transaction's data is sealed until the transaction's execution is scheduled, the processing time per transaction is increased. In particular, critical transactions are negatively affected by increased processing time. In contrast, mitigation schemes that leverage fair ordering and input aggregation approaches can realize real-time functionality.  
We observe that batch-order fair ordering schemes~\cite{DBLP:conf/crypto/Kelkar0GJ20, DBLP:conf/asiapkc/KelkarDK22, DBLP:conf/ccs/KelkarDLJK23, DBLP:conf/aft/Kursawe20} rely on liveness, synchronized clocks among miners, and assume a consistent ordering of transactions as received by the orderers---an assumption that, to date, has not been empirically validated due to the absence of thorough analysis on this subject. 

\begin{table}[tb]
    \centering
    \footnotesize
    \scalebox{0.75}{\begin{tabular}{|c|c|c|c|c|}
        \hline
        & Detection Tool & LF & ETA & Additional Notes \\
        \hline
        \hline
        \parbox[t]{2mm}{\multirow{8}{*}{\rotatebox[origin=c]{90}{Bitcoin}}} & \multirow{2}{3.5cm}{\href{https://mempool.space/}{Mempool Open Source Project}~\cite{mempool_space}} & \multirow{2}{*}{\cmark} & \multirow{2}{*}{\cmark$^1$} & \multirow{2}{5.0cm}{$\nmbObservers=24$; $^1$"Block audits" comparing the expected block with the actual block (\cf~Figure~\ref{fig:mempool_block_audit}) }\\
        & & & &  \\
        \cline{2-5}
        & \multirow{2}{3.5cm}{\href{https://blockstream.info/tx/recent}{Blockstream Explorer}~\cite{blockstream}} & \multirow{2}{*}{\cmark} & \multirow{2}{*}{\cmark} & \multirow{2}{5.0cm}{ Simple log of transactions with ETA} \\
        & & & &  \\
        \cline{2-5}
        &\multirow{2}{3.5cm}{\href{https://btcscan.org/tx/recent}{Btcscan}~\cite{btcscan}} & \multirow{2}{*}{\cmark} & \multirow{2}{*}{\cmark} & \multirow{2}{5.0cm}{ Simple log of transactions with ETA } \\
        & & & &  \\
        \cline{2-5}
        & \multirow{2}{3.5cm}{\href{https://www.blockchain.com/explorer/mempool/btc}{Blockchain.com}~\cite{blockchain_com_btc}} & \multirow{2}{*}{\cmark} & \multirow{2}{*}{\xmark} & \multirow{2}{5.0cm}{ Precise reception times } \\
        & & & &  \\
        \cline{1-5}
        \parbox[t]{2mm}{\multirow{6}{*}{\rotatebox[origin=c]{90}{Ethereum}}} & \multirow{2}{3.5cm}{\href{https://eth.tokenview.io/en/pending}{Tokenview}~\cite{tokenview}} & \multirow{2}{*}{\cmark} & \multirow{2}{*}{\xmark} & \multirow{2}{5.0cm}{ Precise reception times } \\
        & & & &  \\
        \cline{2-5}
        & \multirow{2}{3.5cm}{\href{https://etherscan.io/txsPending}{Etherscan}~\cite{etherscan}} & \multirow{2}{*}{\cmark} & \multirow{2}{*}{\cmark} & \multirow{2}{5.0cm}{ Precise reception times \& estimated transaction confirmation duration } \\
        & & & &  \\
        \cline{2-5}
        & \multirow{2}{3.5cm}{\href{https://www.blockchain.com/explorer/mempool/eth}{Blockchain.com}~\cite{blockchain_com_eth}} & \multirow{2}{*}{\cmark} & \multirow{2}{*}{\xmark} & \multirow{2}{5.0cm}{ Precise reception times } \\
        & & & &  \\
        \hline
    \end{tabular}}
    \caption{Summary of existing ad-hoc detection tools.  Services with the label \emph{LF} provide a live feed of recently received transactions. The label \emph{ETA} denotes that the service provides an expected time for transactions to appear in a block. From the analyzed tools, Mempool.space~\cite{mempool_space} is the only one that publicly disclosed the number of their observers.}
    \label{tab:detection_tools}
\end{table}

\section{Observer Locations}\label{ap:tables}

Table~\ref{tab:server_location} shows the distribution of the complete set of 16 nodes, and Table~\ref{tab:server_location_EUS} shows the locations for the subset of 12 nodes located exclusively in the US and Europe. Note that we report additional results on the reception rates of our observer nodes located exclusively in the US and Europe in Section \ref{ap:EUS_results}.

\begin{table}[]
    \centering
    \begin{tabular}{|c|c|c|}
        \hline
        State/Country & Network zone & \#Nodes \\
        \hline
        \hline
         Virginia, USA & us-east & 2 \\
         Oregon, USA & us-west & 2 \\
         \hdashline
         Germany & eu-central & 4 \\
         \hdashline
         Finland & eu-north & 4 \\
         \hdashline
         Singapore & asia & 4 \\
        \hline
        \cline{3-3}
        \multicolumn{2}{c|}{} & 16 \\
        \cline{3-3}
    \end{tabular}
    \caption{The locations of all our observers.}
    \label{tab:server_location}
\end{table}

\begin{table}[]
    \centering
    \begin{tabular}{|c|c|c|}
        \hline
        State/Country & Network zone & \#Nodes \\
        \hline
        \hline
         Virginia, USA & us-east & 2 \\
         Oregon, USA & us-west & 2 \\
         \hdashline
         Germany & eu-central & 4 \\
         \hdashline
         Finland & eu-north & 4 \\
        \hline
        \cline{3-3}
        \multicolumn{2}{c|}{} & 12 \\
        \cline{3-3}
    \end{tabular}
    \caption{Our subset of 12 observers that are located in the US and Europe exclusively.}
    \label{tab:server_location_EUS}
\end{table}

\section{Proofs} \label{sec:proofs}

\begin{proof}[Proof of \cref{th:generalized_multiple_observers}] 
We start our proof by introducing the two sets $\SObservers_1$ and $\SObservers_2$. We assign all observers to either of these two sets such that $\nmbObservers = |\SObservers_1|+|\SObservers_2|$. That is, all observers in $\SObservers_1$ receive $\transaction_1$ before $\transaction_2$ and all observers in $\SObservers_2$ receive $\transaction_2$ before $\transaction_1$.   
We first determine the number of combinations for sampling \nmbObserversSubset out of \nmbObservers observers for $\SObservers_1$. That is, given \nmbObservers observers, there exist exactly $\binom{\nmbObservers}{\nmbObserversSubset}$ distinct combinations of observers to form $\SObservers_1$ with $|\SObservers_1| = \nmbObserversSubset$. 

Next, we rearrange the above equation using the fact that $\SObservers_1$ and $\SObservers_2$ are disjoint sets:
\begin{align}
    &\pr{\nmbObserversSubset = \bigg|\Bigl\{\observer\in\SObservers\ \big|\ (\receivedBeforeBy{\transaction_1}{\transaction_2}{\observer}) \Bigr\}\bigg| }  \\ 
    & = \binom{\nmbObservers}{\nmbObserversSubset} \cdot \pr{\bigwedge_{\observer \in \SObservers_1} \receivedBeforeBy{\transaction_1}{\transaction_2}{\observer}} \cdot \pr{\bigwedge_{\observer \in \SObservers_2} \receivedAfterBy{\transaction_1}{\transaction_2}{\observer} }    \nonumber
\end{align}

Assuming that $\timeVar_{\observer_i}(\transaction_x)$ for $\forall \observer_i \in \SObservers, \forall x$ are drawn from the same distribution with $\timeVar_{\observer_i}(\transaction_x)$ being independent from $\timeVar_{\observer_j}(\transaction_x)$ for $i \neq j$, it holds: 
\begin{equation}
    \pr{\bigwedge_{\observer \in \SObservers} \receivedBeforeBy{\transaction_1}{\transaction_2}{\observer}} = \pr{\receivedBeforeBy{\transaction_1}{\transaction_2}{\observer}}^{|\SObservers|}
\end{equation}

Finally, applying Equation~(\ref{eq:prob_head_int}) yields:
\begin{equation}
    \pr{\receivedBeforeBy{\transaction_1}{\transaction_2}{\observer}}^{|\SObservers|} = \left( \int_0^\infty \CDF{\timeVar_{\observer}(\transaction_1)}{\timeDiff + \iota} \PDF{\timeVar_{\observer}(\transaction_2)}{\iota} d\iota \right)^{|\SObservers|}
\end{equation}

Combining these observations, the probability that any \nmbObserversSubset out of \nmbObservers observers receive transaction $\transaction_1$ before $\transaction_2$ is given by:  
\begin{align}
        \probGMO_{\nmbObservers}(\nmbObserversSubset) = & Pr\Big[\ \nmbObserversSubset = \bigg|\Bigl\{\observer\in\SObservers\ \big|\ (\receivedBeforeBy{\transaction_1}{\transaction_2}{\observer}) \Bigr\}\bigg| \ \Big] \\
        =& \binom{\nmbObservers}{\nmbObserversSubset} \Biggl( \left( \int_0^\infty \CDF{\timeVar_{\observer}(\transaction_1)}{\timeDiff + \iota} \PDF{\timeVar_{\observer}(\transaction_2)}{\iota} d\iota \right)^{\nmbObserversSubset}  \nonumber \\ 
        & \qquad \cdot \left( 1 - \int_0^\infty \CDF{\timeVar_{\observer}(\transaction_1)}{\timeDiff + \iota} \PDF{\timeVar_{\observer}(\transaction_2)}{\iota} d\iota \right)^{\nmbObservers-\nmbObserversSubset} \Biggr) \nonumber
    \end{align}
\end{proof}

\begin{proof}[Proof of \cref{th:predictive_precision}]
    We rearrange the above probability as follows: 
    \begin{align}
 \pr{\sentBefore{\transaction_1}{\transaction_2} \big| \funcZ } = \bigintssss_{-\infty}^{0} \pr{\timeDiff \ \Big|\  \funcZ }\; d\timeDiff 
    \end{align}
    The probability distribution of \timeDiff conditioned on the event $\probGMO_\nmbObservers(\nmbObserversSubset; \receivedBefore{\transaction_1}{\transaction_2})$, i.e., the event that $\nmbObserversSubset$ out of $\nmbObservers$ received $\transaction_1$ before $\transaction_2$, can be determined by applying Bayes' theorem: 
    \begin{align}
        \pr{\timeDiff \ \Big|\  \funcZ } = \frac{ \pr{\funcZ \ \Big|\ \timeDiff} \cdot \pr{\timeDiff} }{ \bigintss_{-\infty}^{\infty} \left( \pr{\funcZ \ \Big|\ \widehat{\timeDiff}} \cdot \pr{\widehat{\timeDiff}} \right) \; d\widehat{\timeDiff} } 
    \end{align}
    Altogether, it holds: 
    \begin{align}
        \PPV_{\nmbObservers}(\nmbObserversSubset) &= \pr{\sentBefore{\transaction_1}{\transaction_2} \big| \funcZ }  \nonumber \\
        &= \bigintss_{0}^{\infty} \left( \frac{ \pr{\funcZ \ \Big|\ \timeDiff} \cdot \pr{\timeDiff} }{ \bigintss_{-\infty}^{\infty} \pr{\funcZ \ \Big|\ \widehat{\timeDiff}} \cdot \pr{\widehat{\timeDiff}} \; d\widehat{\timeDiff} } \right) \; d\timeDiff 
    \end{align} 
\end{proof}

\begin{proof}[Proof of \cref{th:DAB}]
It holds for any observer \observer:
\begin{align}
    \bigwedge_{i} \receivedBeforeBy{\transaction_i}{\transaction}{\observer} \ \Leftrightarrow \ \timeVar_{\observer}(\transaction) > \censorTime \quad \text{and} \quad \pr{\timeVar_{\observer}(\transaction) > \censorTime} = 1 - \CDF{\timeVar_{\observer}(\transaction)}{\censorTime}
\end{align}

Then, assuming that $\timeVar_{\observer_i}(\transaction_x)$ is independent from $\timeVar_{\observer_j}(\transaction_x)$ for any $i \neq j$, it holds: 
    \begin{align}
        \pr{\bigwedge_{\observer \in \SObservers}\left( \bigwedge_{i} \receivedBeforeBy{\transaction_i}{\transaction}{\observer} \right)} = \pr{\timeVar_{\observer}(\transaction) > \censorTime}^{\nmbObservers} = \Big(1 - \CDF{\timeVar_{\observer}(\transaction)}{\censorTime}\Big)^{n} 
    \end{align}
\end{proof}

\section{Observers located in US \& Europe} \label{ap:EUS_results}

This section reports selected results (Table~\ref{tab:res_reception_general_both_EUS}) on transactions recorded in the Bitcoin and the Ethereum networks using observers located exclusively in the US and Europe (\cf~Table~\ref{tab:server_location_EUS}), mimicking the current geographic distribution of Bitcoin and Ethereum nodes and Mempool.space~\cite{mempool_space} observers. Specifically, we report additional evaluation results using a subset of the transactions evaluated in Section~\ref{sec:empirical_validation_network} filtered by reception entries of nodes located in the US and Europe. 
This additional evaluation is motivated by the fact that, while our main evaluation builds on measurements recorded by observers globally distributed around the world (\cf~Table~\ref{tab:server_location}), some blockchains have a heavy accumulation of nodes in the US and Europe. That is, according to data from \href{https://bitnodes.io/}{bitnodes.io}, \href{https://www.ethernodes.org/countries}{ethernodes.org}, and \href{https://mempool.space/lightning/group/the-mempool-open-source-project}{mempool.space}, the majority of both Bitcoin's and Ethereum's nodes are located in these regions. Nevertheless, our findings reported in Section~\ref{sec:bc_eval} and Section~\ref{sec:eth_eval} remain consistent regardless of whether observers are restricted to the US and Europe or distributed more globally, including regions such as Singapore.

\begin{table}[]
    \centering
    \scalebox{0.82}{\begin{tabular}{|c||r|r|r||r|r|r|}
        \cline{1-7} 
        \multicolumn{7}{|c|}{US \& Europe only} \\
        \cline{1-7} 
        & \multicolumn{3}{c||}{Bitcoin (\numprint{3134824} Txs)} & \multicolumn{3}{c|}{Ethereum (\numprint{6552228} Txs)} \\
        \cline{1-7}
        \multirow{2}{*}{\nmbObserversSubset} & \multicolumn{1}{c|}{Received} & \multicolumn{2}{c||}{$\mathcal{C}_{order}(\nmbObserversSubset)$} & \multicolumn{1}{c|}{Received} & \multicolumn{2}{c|}{$\mathcal{C}_{order}(\nmbObserversSubset)$} \\
         & \multicolumn{1}{c|}{Txs} & \multicolumn{1}{c|}{All Txs} & \multicolumn{1}{c||}{\WRTxs} & \multicolumn{1}{c|}{Txs} & \multicolumn{1}{c|}{All Txs} & \multicolumn{1}{c|}{\WRTxs} \\
        \hline
        \hline
        $\geq6$ nodes  &  99.44\% & 98.89\% & 100.0\% & 96.80\% & 93.48\% & 100.00\%   \\
        $\geq7$ nodes  &  99.38\% & 98.76\% & 100.0\% & 96.55\% & 92.68\% & 99.80\%   \\
        \hdashline 
        $\geq8$ nodes  & 99.25\%  & 98.51\% & 100.0\% & 96.20\% & 91.56\% & 99.39\%    \\
        $\geq9$ nodes  & 99.11\%  & 98.21\% & 100.0\% & 95.95\% & 90.71\% & 99.00\%    \\
        \hdashline 
        $\geq10$ nodes & 98.81\%  & 97.60\% & 100.0\% & 95.65\% & 89.88\% & 98.60\%    \\
        $\geq11$ nodes & 97.99\%  & 95.95\% & 99.99\% & 95.24\% & 88.90\% & 98.21\%    \\
        \hdashline
        12 nodes       & 95.04\%  & 90.32\% & 99.98\% & 87.72\% & 75.27\% & 97.80\%    \\
        \hline
    \end{tabular}}
    \caption{\emph{Received Txs} denotes the number of transactions recorded by at least \nmbObserversSubset out of $\nmbObservers=16$ observers in Bitcoin and Ethereum. \emph{$\mathcal{C}_{order}(\nmbObserversSubset)$} denotes the fraction of transactions received in a consistent order by our 16 nodes. Results labeled with \emph{\WRTxs} are based on the intersection of transactions received by all 16 observers. Results labeled with \emph{All Txs} include all transactions.}
    \label{tab:res_reception_general_both_EUS}
\end{table}

\section{Impact of Measurement Time} \label{ap:first_measurement_results}
In addition to the empirical result reported in Section~\ref{sec:empirical_validation_network}, we provide additional results from an initial measurement with 20 Bitcoin nodes captured in October 2024. The locations of our 20 nodes are highlighted in Table~\ref{tab:server_location_first_measurement}. 

Our observers were similarly hosted by Hetzner cloud~\cite{hetzner_cloud} with an identical setup, i.e., four dedicated vCPU kernels, 16GB RAM, SSD storage, an AMD EPYC-Milan processor, and reachable by an IPv4 address. For software, we used \emph{bitcoind 25.0} as our Bitcoin client. Identically to the evaluation in Sections~\ref{sec:bc_eval}, each observer logged all incoming transactions and their exact reception time, and we compare the transactions' reception times among our observers.

\begin{table}[h]
    \centering
    \scalebox{1.0}{\begin{tabular}{|c|c|c|}
        \hline
        State/Country & Network zone & \#Nodes \\
        \hline
        \hline
         Virginia, USA & us-east & 4 \\
         Oregon, USA & us-west & 4 \\
         \hdashline
         Germany (south) & eu-central & 4 \\
         Germany (east) & eu-central & 4 \\
         \hdashline
         Finland & eu-north & 4 \\
        \hline
        \cline{3-3}
        \multicolumn{2}{c|}{} & 20 \\
        \cline{3-3}
    \end{tabular}}
    \caption{The locations of the 20 nodes from our initial measurement.}
    \label{tab:server_location_first_measurement}
\end{table}

Over a period of 189 hours, we recorded approximately $5.7$ million Bitcoin transactions and $3.4$ million Ethereum transactions over 106 hours. Our results on the Bitcoin measurement are highlighted in Table~\ref{tab:res_reception_general_BC_first_measurement}. It shows that only 94.2\% of all Bitcoin transactions were received by all observers, while 98.21\% reached at least 15, and 98.99\% reached a majority of at least 11 nodes. On average, each node received $98.5\% \pm 0.33\%$ of all transactions. We again determine an empirical estimate of $\probGMO_\nmbObservers(\nmbObserversSubset)$ by sampling \numprint{100000} random pairs of transactions $\bigl(\transaction^{(i)}_0, \transaction^{(i)}_1\bigr)$ with $i \in I:= [1, \numprint{100000}]$, i.e., \numprint{200000} transactions uniform at random from our recorded transaction sets and pair them up randomly. 
We observe that 88.51\% of the randomly drawn transaction pairs from the complete dataset are witnessed by all 20 observers in the same order. For transactions that are received by all observers (\emph{\WRTxs}), we observe that 99.4\% of the sampled pairs were received in consistent order. 

The results in Table~\ref{tab:res_reception_general_BC_first_measurement} show similar trends to those in Table~\ref{tab:res_reception_general_both_LS}, indicating that our findings are robust across different measurement periods.

\begin{table}[tbp]
    \centering
    \scalebox{1.0}{\begin{tabular}{|c||r|r||r|r|}
        \hline
        \multirow{2}{*}{\nmbObserversSubset} & \multicolumn{2}{c||}{Received Txs} & \multicolumn{2}{c|}{$\mathcal{C}_{order}(\nmbObserversSubset)$} \\
         & \multicolumn{1}{c|}{Abs.} & \multicolumn{1}{c||}{Rel.} & \multicolumn{1}{c|}{All Txs} & \multicolumn{1}{c|}{\WRTxs} \\
        \hline
        \hline
        $\geq10$ nodes & \numprint{5658000} & 99.08\% & 98.12\% & 100.0\% \\
        $\geq11$ nodes & \numprint{5652775} & 98.99\% & 97.94\% & 100.0\% \\
        \hdashline
        $\geq12$ nodes & \numprint{5647283} & 98.98\% & 97.73\% & 99.99\% \\
        $\geq13$ nodes & \numprint{5638991} & 98.75\% & 97.43\% & 99.99\% \\
        \hdashline
        $\geq14$ nodes & \numprint{5625313} & 98.51\% & 96.96\% & 99.98\% \\ 
        $\geq15$ nodes & \numprint{5607949} & 98.21\% & 96.36\% & 99.97\% \\
        \hdashline
        $\geq18$ nodes & \numprint{5521894} & 96.70\% & 93.40\% & 99.91\% \\
        \hdashline
        20 nodes       & \numprint{5380042} & 94.21\% & 88.51\% & 99.72\% \\
        \hline
        \hline
        Total & \multicolumn{2}{c||}{\numprint{5710445}} & \multicolumn{2}{c}{} \\
        \cline{1-3}
    \end{tabular}}
    \caption{\emph{Received Txs} denotes the number of transactions recorded by at least \nmbObserversSubset out of $\nmbObservers=20$ observers in the Bitcoin network. \emph{$\mathcal{C}_{order}(\nmbObserversSubset)$} denotes the fraction of transactions received in a consistent order by our 20 nodes. Results labeled with \emph{\WRTxs} are based on the intersection of transactions received by all 20 observers. Results labeled with \emph{All Txs} include all transactions.}
    \label{tab:res_reception_general_BC_first_measurement}
\end{table}

\section{Block Audit}

\begin{figure}[tbp]
    \centering
    \begin{subfigure}[t]{0.45\columnwidth}
	\includegraphics[width=\linewidth]{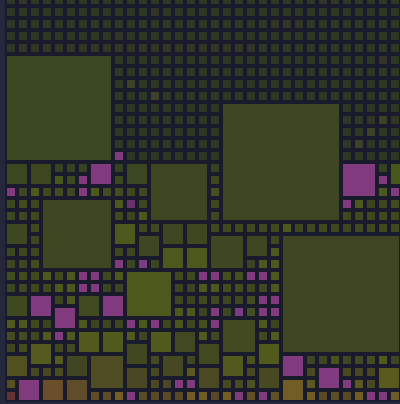}
	\caption{Expected block view. }
	\label{fig:mempool_expected_block}
    \end{subfigure}
    \begin{subfigure}[t]{0.45\columnwidth}
	\includegraphics[width=\linewidth]{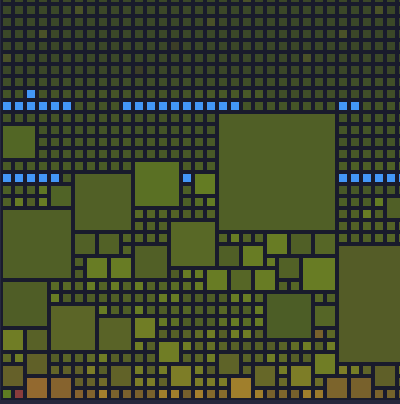}
	\caption{Actual block view.}
	\label{fig:mempool_actual_block}
    \end{subfigure}
    \caption{ An excerpt of the block audit by the Mempool Open Source Project~\cite{mempool_space} for Bitcoin block \numprint{859502}. This example shows the comparison of (a) the expected block and (b) the actual block. Each square denotes a transaction, with the square's color indicating the transaction's label.} 
    \label{fig:mempool_block_audit}
\end{figure}

The Mempool Open Source Project~\cite{mempool_space} provides a preview of a potential block template. This block template incorporates all transactions that are expected to be included in the next block. Once a block is mined, the tool compares the expected block template with the actual block and highlights all differences (\cf~Figure~\ref{fig:mempool_block_audit}). 
In particular, it assesses each transaction according to its reception time and fees. For instance, the tool exposes transactions added to the block template without being received by any of their nodes and transactions with marginal fees. Figure~\ref{fig:mempool_block_audit} depicts a block audit for Bitcoin block \numprint{859502}. Here, blue-colored transactions in \cref{fig:mempool_actual_block} denote transactions that are present in the actual block but not included in the expected block. Pink-colored transactions in \cref{fig:mempool_expected_block} denote transactions that are present in the expected block but not in the actual block.

\end{document}